\documentclass[pra,reprint,aps,notitlepage,nofootinbib,superscriptaddress, preprintnumbers]{revtex4-2}
\usepackage{amsmath}
\usepackage{amssymb}
\usepackage{bm,braket}
\usepackage{qcircuit}
\usepackage{float}
\usepackage{xcolor}
\usepackage{comment}
\usepackage{hyperref}
\usepackage{multirow}
\usepackage{subfig}
\def\be{\begin{equation}}
\def\ee{\end{equation}}
\def\bea{\begin{eqnarray}}
\def\eea{\end{eqnarray}}
\def\bes{\begin{eqnarray}}
\def\ees{\end{eqnarray}}
\def\bi{\begin{itemize}}
	\def\ei{\end{itemize}} 
\usepackage{amsthm}
\theoremstyle{definition}

\definecolor{rindou1}{rgb}{0.4431,0.2862,0.7960}
\definecolor{rindou2}{rgb}{0.0078,0.1215,0.4392}
\definecolor{lapis}{rgb}{0.0.0470,0.2941,0.5568}
\definecolor{mn}{rgb}{0.15, 0.35, 0.95}

\hypersetup{%
	colorlinks,
	citecolor=magenta,
	filecolor=BrickRed,
	linkcolor=rindou2,
	urlcolor=blue,
	linktocpage=true
}

\usepackage{tikz}
\usetikzlibrary{braids}

\usepackage{autoaligne}
\DeclareMathSizes{10}{9}{6}{4}
\usepackage{tabularray}
\definecolor{vio}{RGB}{153,85,255} 
\definecolor{blu}{RGB}{42,127,255} 
\definecolor{olv}{RGB}{170,212,0} 
\definecolor{prp}{RGB}{211,95,188} 

\begin{document}

\title{Mode Distinguishability in Multi-photon Interference}
	\author{Noah Crum}
	\email{ncrum@vols.utk.edu}
	\affiliation{Department of Physics and Astronomy, The University of Tennessee, Knoxville, TN 37996-1200, USA}

\author{Md Mehdi Hassan}
\email{mhassa11@vols.utk.edu}
	\affiliation{Department of Physics and Astronomy, The University of Tennessee, Knoxville, TN 37996-1200, USA}
 
 \author{Adrien Green}
	\email{agreen91@vols.utk.edu}
	\affiliation{Department of Physics and Astronomy, The University of Tennessee, Knoxville, TN 37996-1200, USA}

	\author{George Siopsis}
	\email{siopsis@tennessee.edu}
	\affiliation{Department of Physics and Astronomy, The University of Tennessee, Knoxville, TN 37996-1200, USA}
\date{\today}
\begin{abstract}
The Hong-Ou-Mandel (HOM) effect is a quintessential process in various quantum information technologies and quantum optics applications. In this work, we investigate multi-photon interference, developing a model for the simultaneous characterization of polarization and spectro-temporal mode mismatch on the coincidence probabilities including the effects of realistic imperfections of devices used in HOM experiments. We also study the coincidence probability for coherent states as a function of source intensity, as well as spectro-temporal and polarization mismatch of the incident beams. We apply our model to the case of multi-photon interference from independent sources and analyze the consequences of mode mismatch in various instances that occur in quantum networking including entanglement swapping, quantum key distribution, quantum sensing, quantum optical classification, and photonic quantum computing.
\end{abstract}
\maketitle

\section{Introduction}
Hong, Ou, and Mandel pioneered the interference of two photons under the influence of spatial mixing induced by a beam splitter, known as HOM interference \cite{HOM}. The degree of interference is a function of the indistinguishability in all degrees of freedom of the input photons. The phenomenon is understood through the coincidence rate of detection for photons output by a beam splitter. Indistinguishable photons exhibit a characteristic dip in their coincidence rate whose severity relates to the degree of indistinguishability of the inputs. The original HOM experiment sought to characterize the temporal distinguishability of single photons via fine control and measurement of photon bandwidths and timing intervals. The interference of single photons with different types of distinguishability in different photonic modes is covered extensively in \cite{brańczyk2017hongoumandel}.

The nature of HOM interference provides numerous opportunities for practical applications to the problem of clock synchronization \cite{Giovannetti_QCS}, quantum teleportation\cite{bouwmeester_experimental_1997}, quantum computation \cite{Knill2001, Kok_2007, O'Brien2003}, non-classical communication \cite{Lo_2012, Xu_2020}, and quantum-enhanced metrology \cite{Ulanov2016}. It is also widely used in the areas of boson sampling \cite{Crespi2013}, quantum random walks \cite{Bromberg_2009}, and quantum dense coding \cite{Mattle_1996}.

Theoretical models of quantum interference are fundamental to the operation of quantum networking protocols. However, in practical scenarios, real-world imperfections, arising from source, channel, and detector mismatches, complicate the realization of ideal interference. This paper extends the theoretical framework of HOM interference to incorporate polarization and spectro-temporal mode mismatch, addressing a pressing issue in the deployment of quantum networks composed of heterogeneous quantum technologies.

In heterogeneous networks, nodes often rely on distinct quantum resources and photonic sources, each with varying spectral, temporal, and polarization characteristics. For example, spontaneous parametric down-conversion (SPDC) sources \cite{Rarity_2005, Pittman_2003}, quantum dots \cite{Patel2010}, single atoms \cite{Vural_2018}, trapped ions \cite{Gerber_2009, Toyoda2015}, and nitrogen-vacancy centers in diamond \cite{NVC} generate photons with markedly different spectral properties, making precise mode matching across nodes a significant challenge. By simultaneously accounting for polarization and spectral mismatches, this work advances techniques for modeling and mitigating these practical issues.

In practical quantum networking scenarios, HOM interference visibility is a critical parameter tied to fundamental metrics such as the fidelity of entanglement swapping \cite{Ent_swap_bennet, Jin2015}, the quantum bit error rate (QBER) in Measurement Device Independent Quantum Key Distribution (MDI-QKD) \cite{Lo_2012}, and the success probability of Optical Bell State Measurements (OBSM). These metrics underpin the performance and scalability of quantum communication systems.

HOM interference plays a central role in quantum optical classifiers \cite{roncallo2024quantum} which are emerging computational models that leverage quantum optical systems to perform classification tasks more efficiently than classical counterparts. These classifiers utilize the quantum properties of photons, such as superposition and interference, to process information. HOM interference ensures that photons, typically representing data points or decision boundaries, interfere in a way that leads to a meaningful classification outcome. High visibility in HOM interference ensures that the photons are indistinguishable, which is crucial for reliable and accurate classification. Any distinguishability, such as spectral or polarization mismatches, can reduce the fidelity of the classifier's output, making the optimization of mode matching a key factor in the performance of quantum optical classifiers. 

HOM interference facilitates two-photon interactions in photonic quantum computing \cite{alexander2024manufacturable}, which relies on manipulating the quantum states of photons to perform computational tasks, often utilizing linear optical elements such as beam splitters and phase shifters. When photons interfere destructively, they can be used to implement key quantum logic gates, such as controlled-NOT (CNOT) gates, a necessary building block of quantum algorithms. In linear optical quantum computing, the ability to manipulate and control the interference patterns of photons, ensured by high-quality HOM interference, is essential for creating fault-tolerant quantum circuits. Imperfections in interference, caused by mode distinguishability, can lead to errors in computation, emphasizing the importance of managing spectral and polarization modes in these systems.

The multi-mode, multi-photon framework developed in this work provides a comprehensive model to quantify HOM interference visibility under realistic experimental conditions. Realistic imperfections—including beam splitter asymmetries, input intensity mismatches, and polarization and spectro-temporal mode mismatches—degrade interference visibility \cite{Wang_Device_imperfections, Moschandreou_2018}. Previous studies have addressed specific contributions to this degradation. For example, Moschandreou \textit{et al.}\ \cite{Moschandreou_2018} explored the impact of input intensity on visibility, while Pradana \textit{et al.}\ \cite{Pradana_2019} introduced a Gaussian spectral model for coherent states. Building on these works, we extend the analysis to simultaneously consider polarization mismatch, spectro-temporal overlap, input intensities, beam splitter asymmetries, and detector efficiencies. 

This study has broad implications for the design and optimization of quantum networking protocols. Specifically, the theoretical insights presented here pave the way for:

\begin{enumerate}
    \item \textbf{Improved Fidelity in Entanglement Distribution}: Predicting and managing mode mismatches facilitates the design of spectral filtering and active polarization control strategies, essential for robust entanglement swapping.
    
    \item \textbf{Enhanced Performance in Quantum Key Distribution}: HOM interference visibility directly impacts the QBER in MDI-QKD \cite{Rubenok_TPI_QKD, Liu_exp_MDI, Tang_exp_pol_encode_MDI, Tang_MDI_met_net, Tang_exp_MDI, Comandar2016, Reaz_2024, Yuan_short_pulse_interference, Comandar2016, FerreiradaSilva:15}. By identifying operational regimes for optimal visibility, this work minimizes security vulnerabilities.
    
    \item \textbf{Scalability in Quantum Networks}: Heterogeneous nodes demand adaptive protocols. The framework discussed here supports the integration of disparate technologies, enabling cohesive and scalable quantum networks.
\end{enumerate}

 Ultimately, this work bridges the gap between theoretical models and practical implementation, offering a path forward for realizing high-performance quantum networks. By addressing the challenges posed by heterogeneous quantum technologies, this study supports the next generation of quantum communication and computation systems.

The structure of this paper is as follows. In Section \ref{sec:2}, we introduce the multi-photon state featuring spectro-temporal, polarization, and input spatial modes. Next, we present the general expression for the coincidence probability in the HOM setup as a function of photon number, polarization mismatch, and spectro-temporal overlap. 
Then, we consider the case of generally asymmetric quantum channels on the HOM inputs. In Section \ref{sec:coherent}, we extend the multi-photon model to the case of coherent states and characterize the coincidence probability in this instance. In Section \ref{sec:5}, we discuss applications to entanglement swapping \cite{Ent_swap_bennet, Jin2015}, MDI-QKD, a quantum optical classifier \cite{roncallo2024quantum}, and photonic quantum computing \cite{alexander2024manufacturable}. Finally, in Section \ref{sec:6}, we present our conclusions. In Appendix \ref{app:4}, we discuss detector efficiency functions. Details of calculations can be found in Appendix \ref{app:1}.  Appendix \ref{app:3} contains examples of commonly used spectral envelopes. In Appendix \ref{app:Quant_channels}, we discuss quantum channels. 

\section{Multi-photon States}\label{sec:2}

In this Section, we develop the theoretical framework for understanding the interference effects and detection probabilities in systems of multiple photons. The normalized multi-photon wavefunction is introduced for a set of indistinguishable photons, taking into account both spectral and polarization modes. The operator formalism for polarization states is defined in terms of horizontal ($H$) and vertical ($V$) directions, allowing us to describe multi-photon states in different polarization modes.
We then analyze how these states evolve through a beam splitter, and compute the coincidence detection probability, which occurs when photons are detected at both output modes, incorporating parameters such as polarization mismatch, spectral overlap, and detector efficiencies.
Special attention is given to the case where the photons from different modes have a time delay between them. Finally, we derive the interference visibility, a key measure for evaluating the performance of the interferometer, which depends on several factors, including detector efficiency, beam splitter properties, and the spectral and polarization overlap of the input photons.

\subsection{Theory}

The normalized multi-photon wavefunction for $N$ indistinguishable photons in a single wave packet (i.e., all photons have identical spectral function $\phi(\omega)$) is given by
\begin{equation}
    \ket{N, \phi} = \frac{1}{\sqrt{N!}} (\hat{A}^\dagger[\phi])^N \ket{0}
\end{equation}
where $\hat{A}^\dagger[\phi] = \int d\omega\, \phi(\omega) \hat{a}^\dagger (\omega)$, and $\int d\omega |\phi(\omega)|^2=1$ \cite{Pradana_2019}. To account for polarization of the state, we introduce a pair of orthogonal directions: horizontal ($H$) and vertical ($V$). The polarization of the $N$-photon state in spatial mode $A$ is $\hat{\epsilon}_A$, expressed in terms of the chosen directions $\hat{\epsilon}_{H,V}$ gives rise to the operator
\begin{equation}\label{eq:2a}
    \hat{A}^\dagger[\hat{\epsilon}_A,\phi] = (\hat{\epsilon}_A \cdot \hat{\epsilon}_H) \hat{A}_H^\dagger [\phi] + (\hat{\epsilon}_A \cdot \hat{\epsilon}_V)\hat{A}_V^\dagger [\phi]\ , 
\end{equation}
where $\hat{A}_i^\dagger[\phi] = \int d\omega\, \phi(\omega) \hat{a}_i^\dagger (\omega)$, $i\in \{ H,V \}$.

The HOM setup involves two input beams, labeled $A$ and $B$, and created with $\hat{a}_i^\dagger$ and $\hat{b}_i^\dagger$, respectively ($i\in \{ H,V \}$). The two input spatial modes mix at the beam splitter leading to the following transformation, $\hat{U}_{BS}$, of creation operators,
\begin{equation}
\hat{U}_{BS}\ : \ \begin{array}{ll}
    \hat{a}_i^\dagger &\rightarrow c_i^\dagger = t \hat{a}_i^\dagger + r \hat{b}_i^\dagger \\
    \hat{b}_i^\dagger &\rightarrow d_i^\dagger = r \hat{a}_i^\dagger - t \hat{b}_i^\dagger
\end{array}
\end{equation}
for a beam splitter with reflectivity and transmissivity, $R=r^2$ and $T=t^2$, respectively, where $R+T=1$, ensuring the unitarity of the transformation. 

Consider input states of $m$ photons in mode $A$ with spectral function $\phi_A(\omega)$ and polarization $\hat{\epsilon}_A$ and $n$ photons in mode $B$ with spectral function $\phi_B(\omega)$ and polarization $\hat{\epsilon}_B$. After passing through a T/R beam splitter, we obtain the output state
\begin{align} \label{eq:6}
        \ket{\mathrm{out}} &= \hat{U}_{BS} \left( \ket{m,\hat{\epsilon}_A, \phi_A}_A \otimes \ket{n, \hat{\epsilon}_B, \phi_B}_B \right) \nonumber\\
                         &= \frac{1}{\sqrt{m!n!}} \left(  \hat{C}^\dagger[\hat{\epsilon}_A, \phi_A] \right)^m \left(  \hat{D}^\dagger[\hat{\epsilon}_B, \phi_B] \right)^n \ket{0} 
\end{align}

We are interested in the probability of coincidence detection after passing through the beam splitter, i.e., mutual detection of photons at both output modes of the beam splitter. Evidently, if all of the photons exit the beam splitter at the same output mode, no coincidence can be detected. Therefore, the probability of coincidence detection, $P_{m,n}^{\text{Co}}$, when $m$ photons enter at $A$ and $n$ photons at $B$ given $\Delta_{A(B)}(\hat{\epsilon}_A, \hat{\epsilon}_B, m, n)$, the detector efficiency function of detector $A\ (B)$  (see Appendix \ref{app:4} for details), is 
\begin{equation} \label{eq:7}
    P_{m,n}^{\text{Co}} = 
        \Delta_A \Delta_B-\Delta_AP_{m,n}^{\text{det}}(m+n, 0) - \Delta_BP_{m,n}^{\text{det}}(0,m+n) \ , 
\end{equation}
for $m+n \geq 1$,
where $P_{m,n}^{\text{det}}(m',n')$ is the probability of detecting $m'$ photons at $A$ and $n'$ photons at $B$. The probability of detecting all the photons at one output mode is given by (see Appendix \ref{app:1} for details)
\begin{equation}\label{eq:7a}
P_{m,n}^{\text{det}}(m+n, 0) = T^m R^n  \mathcal{P} \ , \ \ P_{m,n}^{\text{det}}(0,m+n) = T^n R^m  \mathcal{P} 
\end{equation} 
where
\begin{equation} \mathcal{P} = \sum_{j=0}^{\min(m,n)} \binom{m}{j}\binom{n}{j} \left[ \cos\Phi \cos\Theta \right]^{2j}
\end{equation} 
and
\begin{equation}
    \cos\Phi = |\hat{\epsilon}_A \cdot \hat{\epsilon}_B^*| \ , \ \ \cos\Theta = \left| \int d\omega \phi_A^* (\omega) \phi_B(\omega) \right|
\end{equation}
The angles $\Phi$ and $\Theta$ provide a measure of the mismatch in polarization and spectral profiles, respectively, between photons at $A$ and $B$. We deduce the coincidence probability for $m+n\ge 1$,
\begin{equation} \label{eq:9}
    P_{m,n}^{\text{Co}} = 
        \Delta_A \Delta_B-(T^m R^n \Delta_A + T^n R^m \Delta_B) \mathcal{P} 
\end{equation}
Of particular interest is the case in which the two modes have the same spectrum profile but shifted in time by a constant $\tau$, i.e.,
\begin{equation}\label{eq:delay}
    \phi_B (\omega) = \phi_{A} (\omega) e^{i\omega\tau}
\end{equation}
In this case, a useful quantity for understanding the dip in detection probability is the interference visibility in the temporal domain,
\begin{equation} \label{eq:visibility}
    \mathcal{V} = \frac{P^\text{Co}(\tau \rightarrow \pm \infty) - P^\text{Co}(\tau =0)}{P^\text{Co}(\tau \rightarrow \pm \infty)},
\end{equation}
which quantifies the relationship between the point of maximal temporal distinguishability ($\tau \rightarrow \pm \infty$) ,  the instance of no interference, with the point of minimal temporal distinguishability ($\tau =0$). The maximum possible visibility is 1, indicative of high interference. The visibility here is related to the physical parameters of our interferometer such as detector efficiency and beam splitter asymmetry, as well as the temporal, spectral, and polarization properties of the input photons. 

\begin{figure*}[ht!]
    \centering
    \includegraphics[scale=0.29]{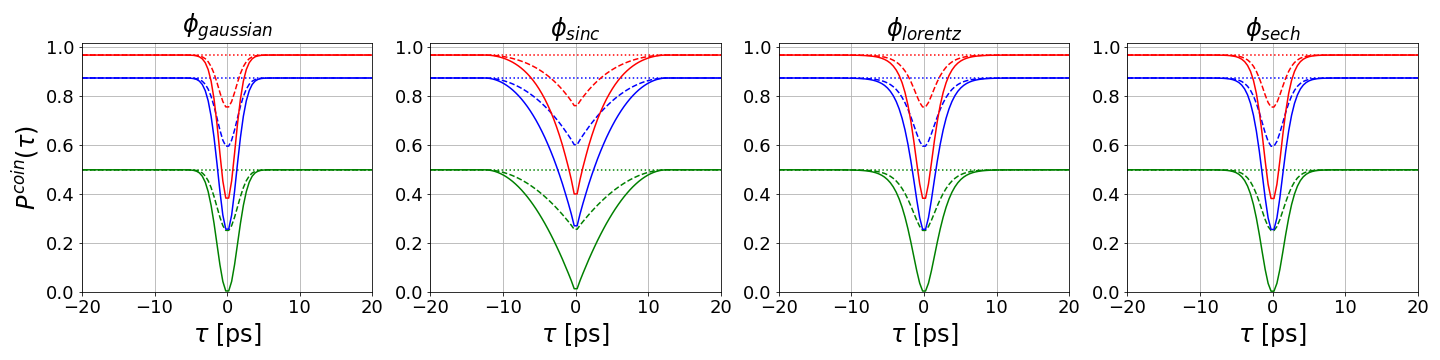}
    \caption{The simulated coincidence probability for each spectral shape under consideration for several values of input photon number $m=n=1,2,3$ for green, blue, and red curves respectively. Each curve is plotted for a separate value of polarization mismatch, $\Phi = 0, \pi/4,$ and $\pi/2$ for solid, dashed, and dotted lines respectively.}
    \label{fig:HOM_spec_matched}
\end{figure*}

\subsection{Numerical Results}
For our numerical study, we are motivated by considerations of HOM for telecom photons with wavelength around 1550 nm, but the presented model can be utilized more generally. We consider spectral bandwidths from tens to a thousand GHz as is typical of many SPDC sources and supports telecom operations in the C-band. We begin our numerical investigation of multi-photon interference with multi-mode distinguishability by interrogating the coincidence probability in the temporal degree of freedom for various spectral profile shapes; namely, Gaussian, sinc, Lorentz, and hyperbolic secant shaped photons. For details on each of these spectral profiles, see Appendix \ref{app:3}. The first study takes spectrally matched input photons, i.e., photons sourced from $A$ and $B$ have the same spectral properties: central frequency, spectral bandwidth, and profile shape but photons in mode $B$ vary in their relative arrival time, $\tau$. Figure \ref{fig:HOM_spec_matched} shows the HOM interference dip in coincidence probability as a function of time delay for the four choices of spectral profile shape. The coincidence probability is altered by the other mode properties of the photons. As seen in Figure \ref{fig:HOM_spec_matched}, increasing the number of input photons increases the probability of coincident detection but reduces the relative depth of the dip. Furthermore, the introduction of polarization distinguishability severely diminishes the strength of the interference.

\begin{figure}[ht!]
    \centering
    \includegraphics[scale=0.3]{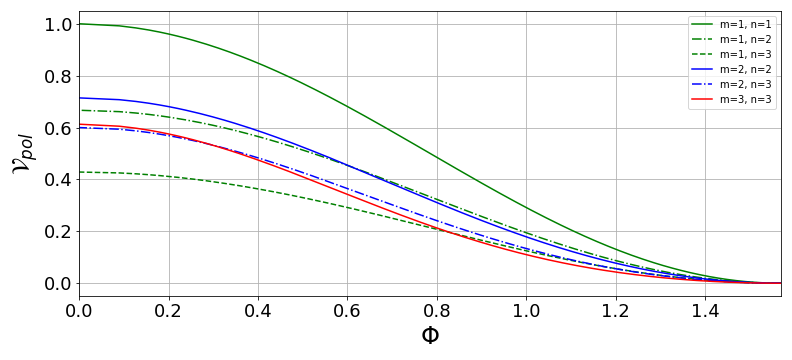}
    \caption{The simulated HOM visibility as a function of polarization mismatch between photons input in input modes $A$ from those in input mode $B$ for several values of input photon number $m,n=1,2,3$. The instances when equal photon numbers are input on the beamsplitter are plotted with solid lines. The dashed and dot-dashed lines indicate instances in which a mismatch in the photon number occurs.}
    \label{fig:vis_vs_Phi}
\end{figure}

Figure \ref{fig:vis_vs_Phi} depicts the interference visibility for different input photon numbers, from $1$ to $3$ as a function of the polarization mismatch of the inputs. In this instance, spectro-temporal indistinguishability is assumed ($\Theta =0$) and the interference visibility is only degraded by the increase in photon number and polarization mismatch, a consequence of the increase in input distinguishability. Of key interest are the instances in which the higher total photon number results in higher visibility than in the mismatched photon number cases.  For instance, the $m=n=2$ has a higher visibility at lower polarization mismatch than the $m=1$, $n=2$ case. 

\begin{figure*}[ht!]
    \centering
    \includegraphics[scale=0.29]{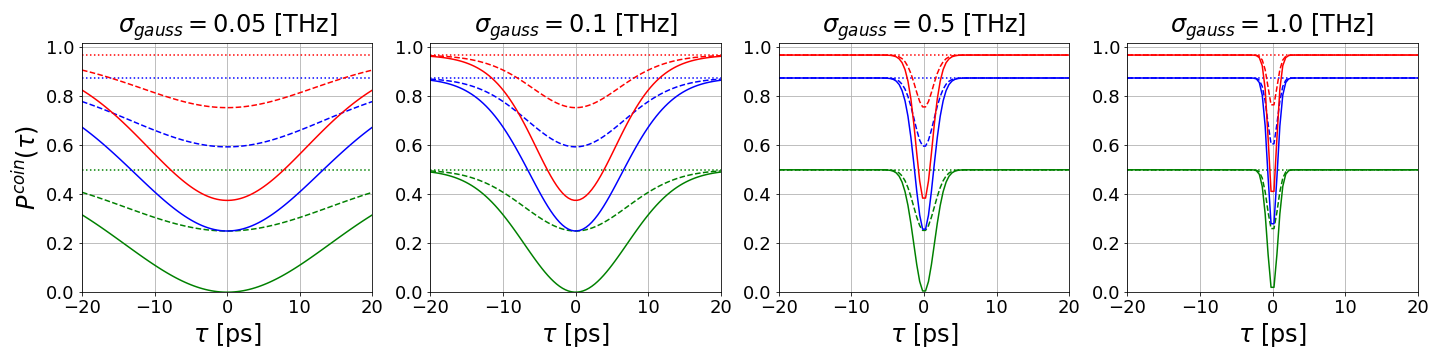}
    \caption{The simulated coincidence probability for spectrally-matched Gaussian photons for various spectral bandwidths (from left to right $\sigma = 0.05,\ 0.10,\ 0.50,\ 1.0 $ [THz]) for several values of input photon number $m=n=1,2,3$ for green, blue, and red curves respectively. Each curve is plotted for a separate value of polarization mismatch, $\Phi = 0, \pi/4,$ and $\pi/2$ for solid, dashed, and dotted line respectively.}
    \label{fig:HOM_spec_matched_var_sig}
\end{figure*}

 Focusing for now solely on Gaussian spectral profiles, Figure \ref{fig:HOM_spec_matched_var_sig} shows how the spectral bandwidth of the input states alters the appearance of the HOM dip. Here, the input photons have the same spectral properties, but the value of the spectral bandwidth is varied. Narrow pulses in the frequency domain correspond to wider pulses in the temporal domain; thus, increases in the spectral bandwidth restrict the temporal width of the HOM dip. The dip is again shown for multiple values off input photons and polarization mismatches. 

\begin{figure*}[ht!]
    \centering
    \includegraphics[scale=0.29]{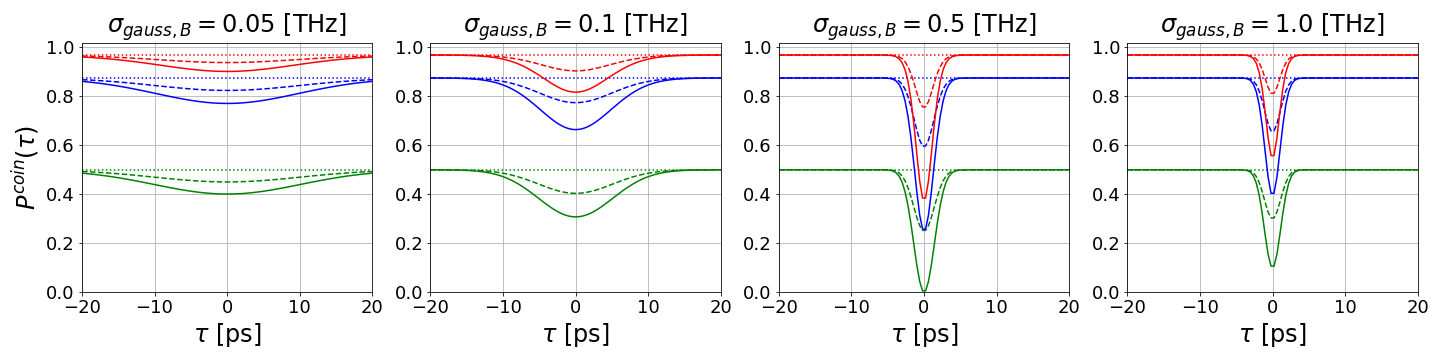}
    \caption{The simulated coincidence probability for photons with allowed mismatch in spectral width for Gaussian photons of various values of spectral bandwidth (from left to right $\sigma = 0.05,\ 0.10,\ 0.50,\ 1.0$ [THz]) for photons input in mode B, while a photon(s) input in mode A has(have) a spectral width parameter $\sigma_A=0.5$ [THz]. Several values of input photon number $m=n=1,2,3$ for green, blue, and red curves respectively are shown and each curve is plotted for a separate value of polarization mismatch, $\Phi = 0, \pi/4,$ and $\pi/2$ for solid, dashed, and dotted line respectively.}
    \label{fig:HOM_var_spec}
\end{figure*}

 Figure \ref{fig:HOM_var_spec} follows similarly to Figure \ref{fig:HOM_spec_matched_var_sig} but now the spectral width of a photon(s) input in $A$ is fixed to $\sigma=0.5$ [THz]. Due to the introduction of a spectral mismatch, the visibility of the interference is reduced. Thus, spectral mismatch strongly affects quantum interference in HOM.

\begin{figure}[h]
    \centering
    \includegraphics[scale=0.3]{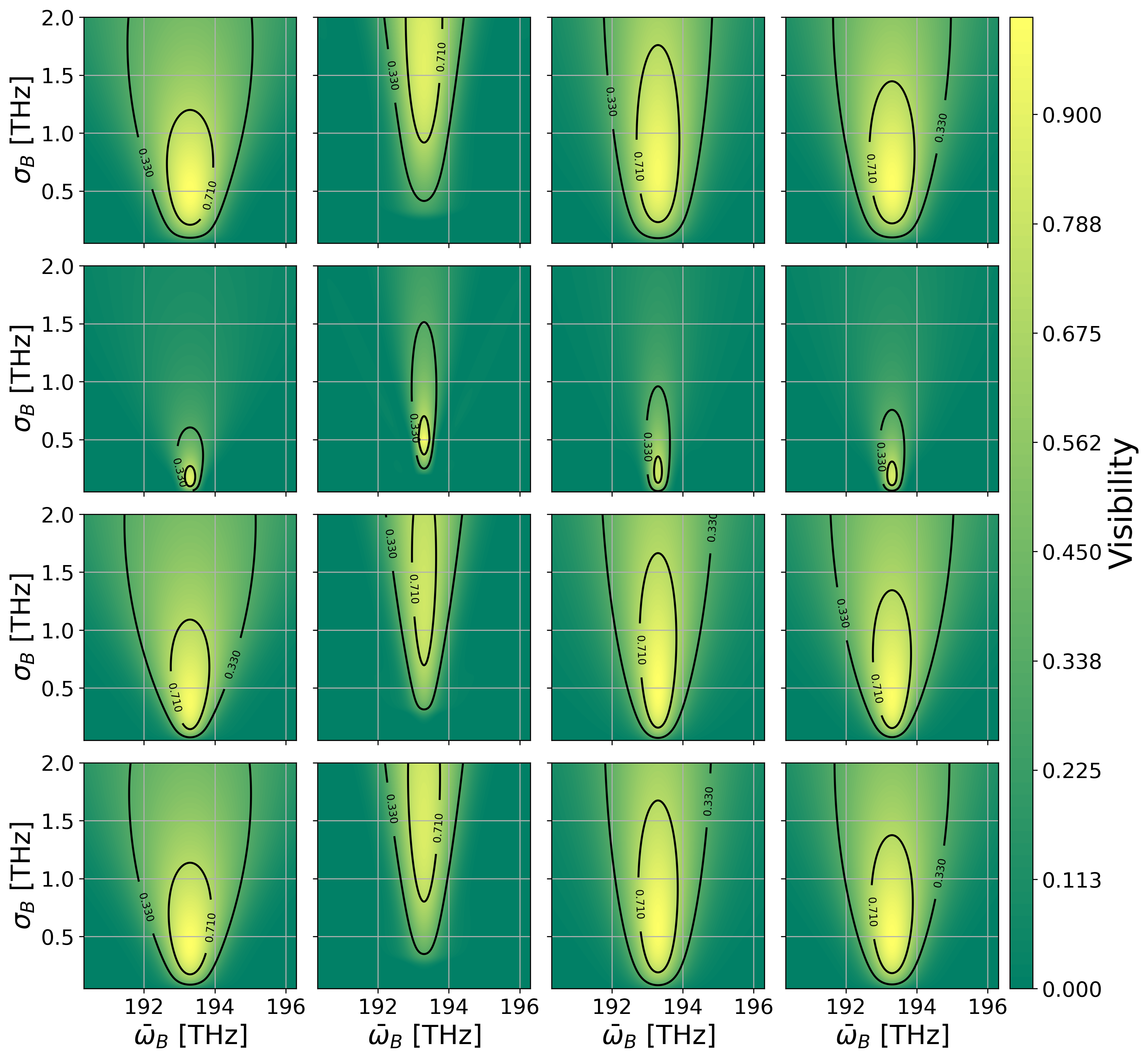}
    \caption{The contour plots depicts the simulated visibility of the HOM interference as a function of the spectral properties of the photon input in B. The spectral properties of the photon input at $A$ are fixed with $\bar{\omega}=193.55$ THz and spectral width of 1 nm. The input spectral profiles across and down are, in order, Gaussian, Sinc, Lorentz, and Sech shaped.}
    \label{fig:m_n_11}
\end{figure}

\begin{figure}[ht!]
    \centering
    \includegraphics[scale=0.3]{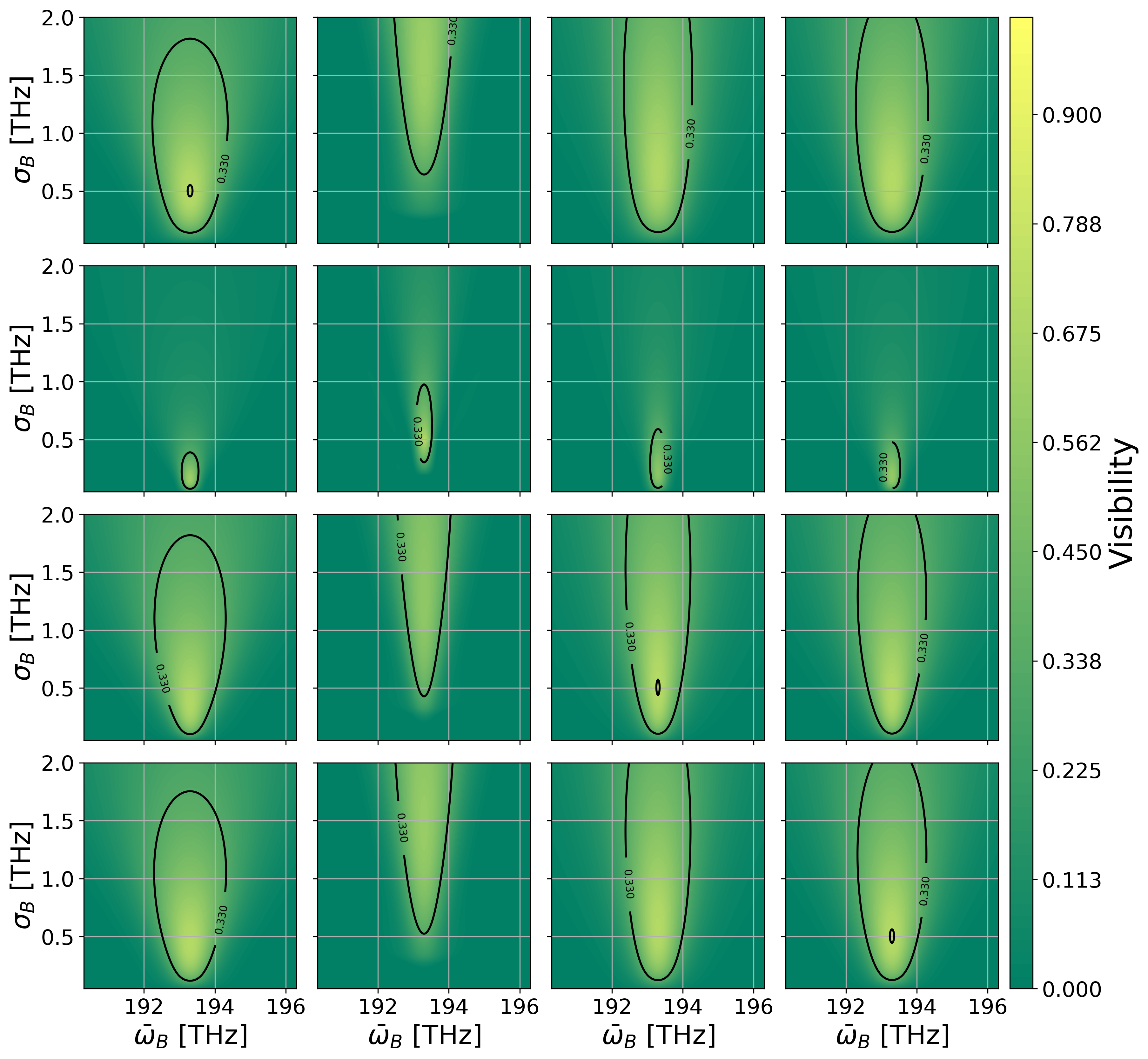}
    \caption{Two photons input at locations $A$ and $B$. The contour plots depicts the  simulated visibility of the HOM interference as a function of the spectral properties of the photons input in B. The spectral properties of the photons input at $A$ are fixed with $\bar{\omega}=193.55$ THz and spectral width of 1 nm. The input spectral profiles across and down are, in order, Gaussian, Sinc, Lorentz, and Sech shaped.}
    \label{fig:m_n_22}
\end{figure}

 To investigate the nature of mismatches in profile shape, Figure \ref{fig:m_n_11} depicts a contour plot of the HOM visibility as a function of the spectral properties of photons from input mode $B$. The figure shows interference for single-photons input in both modes with no polarization mismatch ($\Phi=0$). Despite similar profile shapes, the contours reveal the sensitivity of HOM interference to the spectral properties of the input states. The presence of side-lobes in the sinc-shaped spectral profile yields a much higher sensitivity to the spectro-temporal shape in the interference visibility highlighting the need for active spectral control such as bandwidth filtering or corrective optical elements. Table \ref{tab:m=n=1_vis_and_FWHM_ratio} reports the maximum visibility for each scenario and the ratio of full width at half maximum (FWHM) of the input spectral profiles. Figure \ref{fig:m_n_22} and Table \ref{tab:m=n=2_vis_and_FWHM_ratio} show similar results for sourcing 2 photons in each input mode.

\begin{table}[h]
    \centering
    \begin{tabular}{|c|c|c|c|c|}
        \hline
        \textbf{} & \textbf{$\phi_{\text{gauss}}$} & \textbf{$\phi_{\text{sinc}}$} & \textbf{$\phi_{\text{lorentz}}$} & \textbf{$\phi_{\text{sech}}$} \\
        \hline
        $\phi_{\text{gauss}}$ & 1.00 | 1.00 & 0.89 | 0.12 & 0.97 | 0.90 & 0.99 | 0.79 \\
        \hline
        $\phi_{\text{sinc}}$ & 0.88 | 8.10 & 0.99 | 1.00 & 0.80 | 7.39 & 0.85 | 6.84 \\
        \hline
        $\phi_{\text{lorentz}}$ & 0.97 | 1.10 &  0.81 | 0.13 & 1.00 | 1.00 & 0.99 | 0.88 \\
        \hline
        $\phi_{\text{sech}}$ & 0.99 | 1.27 & 0.86 | 0.14 & 0.99 | 1.14 & 1.00 | 1.00 \\
        \hline
    \end{tabular}
    \caption{Numerical results for max visibility for two-photon interference for each pairing of spectral shape. The columns correspond to the spectral profile of the photon input at A while the rows indicate the spectral shape of photons input at B. The entries report the maximum possible visibility and the ratio of FWHM of photon A to that of photon B. Disparities in the shape of the sourced spectral profiles lead to a reduction in the HOM visibility. }
    \label{tab:m=n=1_vis_and_FWHM_ratio}
\end{table}

\begin{table}[ht]
    \centering
    \begin{tabular}{|c|c|c|c|c|}
        \hline
        \textbf{} & \textbf{$\phi_{\text{gauss}}$} & \textbf{$\phi_{\text{sinc}}$} & \textbf{$\phi_{\text{lorentz}}$} & \textbf{$\phi_{\text{sech}}$} \\
        \hline
        $\phi_{\text{gauss}}$ & 0.71 | 1.00 & 0.62 | 0.12 & 0.69 | 0.90 & 0.70 | 0.79 \\
        \hline
        $\phi_{\text{sinc}}$ & 0.61 | 8.10 & 0.70 | 1.00 & 0.54 | 7.39 & 0.58 | 6.49 \\
        \hline
        $\phi_{\text{lorentz}}$ & 0.69 | 1.10 &  0.56 | 0.13 & 0.71 | 1.00 & 0.71 | 0.86 \\
        \hline
        $\phi_{\text{sech}}$ & 0.71 | 1.27 & 0.60 | 0.14 & 0.71 | 1.14 & 0.71 | 1.00 \\
        \hline
    \end{tabular}
    \caption{Numerical results for max visibility of  interference for two-photons in each input for each pairing of spectral shape. The columns correspond to the spectral profile of the photon input at A while the rows indicate the spectral shape of photons input at B. The entries report the maximum possible visibility and the ratio of FWHM of photon A to that of photon B. The increase in photon number reduces visibility but leaves the ratio of FWHM for maximal interference unchanged}
    \label{tab:m=n=2_vis_and_FWHM_ratio}
\end{table}

\subsection{Multi-photon HOM with Quantum Channel}
 In the context of quantum networking, photons are typically sourced from remote locations and traverse a quantum channel before they are utilized in a quantum information processing task or network protocol. See Appendix \ref{app:Quant_channels} for how we describe the formalism for modeling quantum channels acting on a pure photon number state with polarization and a continuum of spectral modes. We consider three types of quantum channels: amplitude damping, depolarizing, and spectral broadening. Each channel is described using Kraus operators to represent the effect on the input state. Each channel affects each mode variety independently. 

 If $\mathcal{E}_{\text{AD}},\ \mathcal{E}_{\text{DP}},\text{ and } \mathcal{E}_{\text{SB}}$ represent the amplitude damping, depolarizing, and spectral broadening channels respectively, the composed channel $\mathcal{E}_{\text{total}}$ is
 \begin{equation}
     \mathcal{E}_{\text{total}} = \mathcal{E}_{\text{SB}} \circ \mathcal{E}_{\text{DP}} \circ \mathcal{E}_{\text{AD}}.
 \end{equation} 
 The action on the input state $\rho$ is 
 \begin{equation}
     \mathcal{E}_{\text{total}}(\rho) = \mathcal{E}_{\text{SB}}(\mathcal{E}_{\text{DP}}(\mathcal{E}_{\text{AD}}(\rho))).
 \end{equation}

\begin{figure}[h]
\centering
\includegraphics[width=\linewidth]{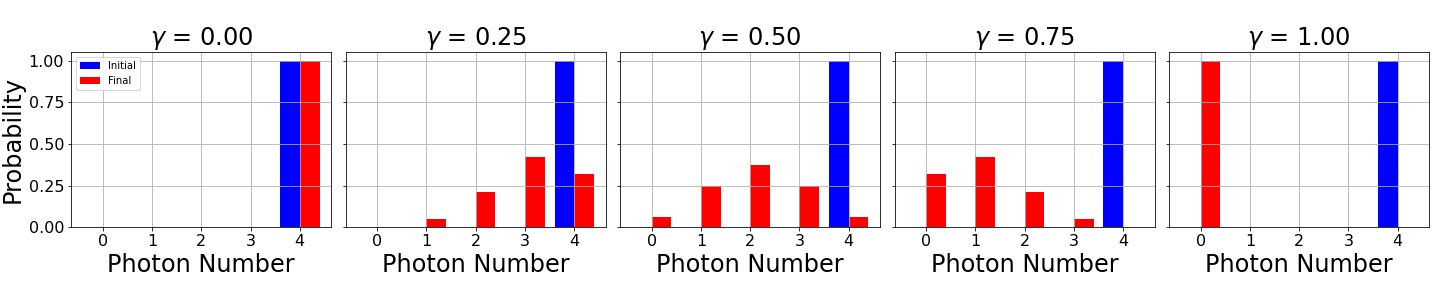}
\caption{Photon number distribution of a 4-photon state for the initial (blue) and post-quantum channel states (red) for increasing values of the amplitude damping parameter $\gamma$.}
\label{fig:amplitude_damping_number_dist}
\end{figure}

\begin{figure}[h]
\centering
\includegraphics[width=\linewidth]{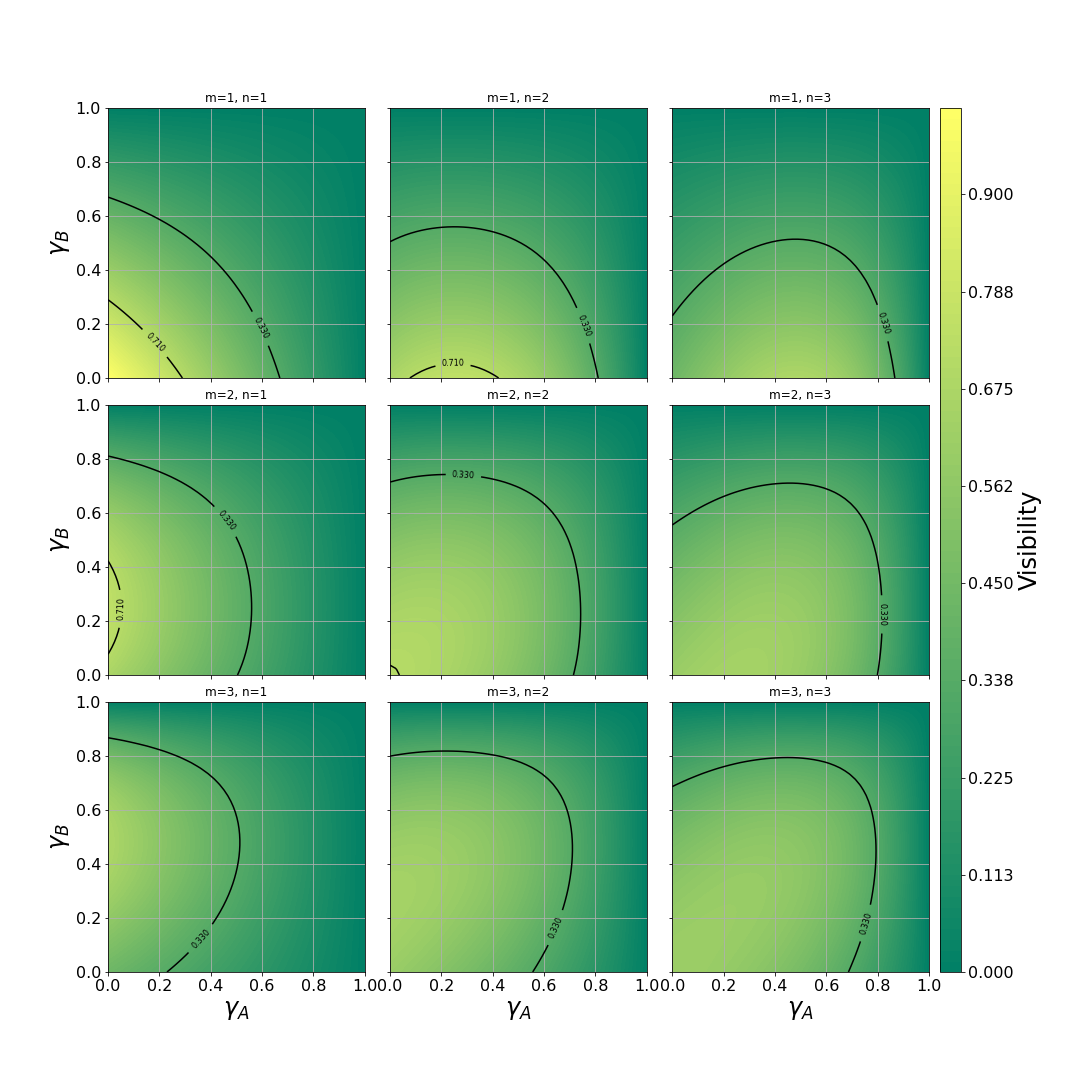}
\caption{Contour plot showing simulated HOM visibility as a function of $\gamma_A$ and $\gamma_B$ for different photon number inputs $m$ and $n$ to the channel. Increasing the damping induced by the channel can yield improved visibility when mitigating a photon number mismatch. }
\label{fig:amplitude_damping_contour}
\end{figure}

Firstly, we simulate the consequence of photon loss in both Alice's and Bob's quantum channels. Figure \ref{fig:amplitude_damping_number_dist} shows how the amplitude damping parameters alter the photon number distribution for a 4-photon Fock state. An increase in the damping parameter directly reduces the single-photon HOM visibility, but when higher photon number states are utilized, the attenuation provided by the quantum channel can alleviate the photon number mismatch at the input, slightly improving the visibility seen in Figure \ref{fig:amplitude_damping_contour}.

\begin{figure}[h]
\centering
\includegraphics[width=\linewidth]{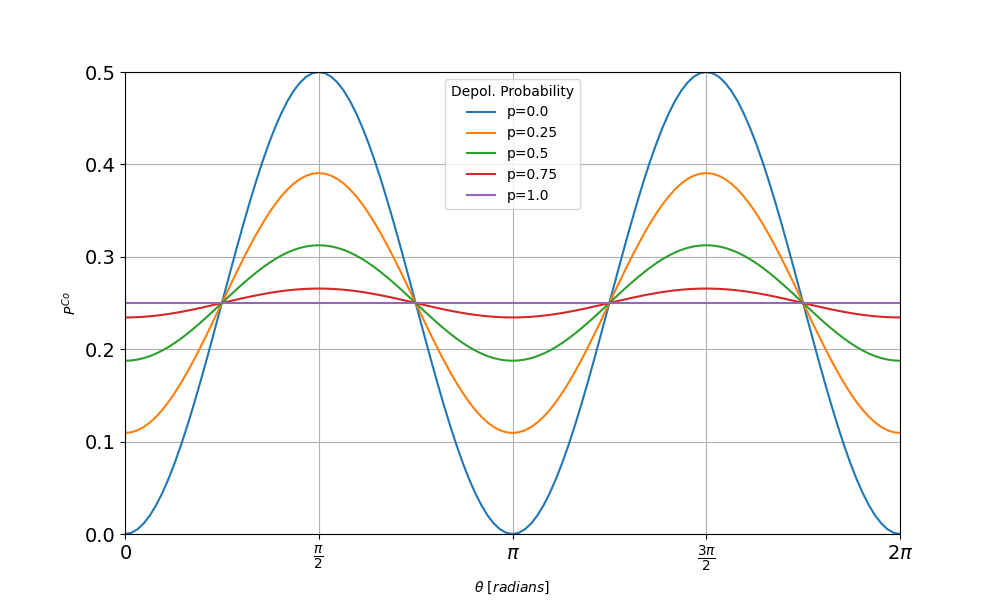}
\caption{Simulated coincidence probabilities for input of single photons subject to depolarization before interference. Photons in one mode have fixed initial polarization prior to the depolarization while the initial polarization of the other input photon is rotated through an angle $\theta$. Each color curve represents the strength of the applied depolarization to both input photons.}
\label{fig:depolarizing_single_photons}
\end{figure}

\begin{figure}[h]
\centering
\includegraphics[width=\linewidth]{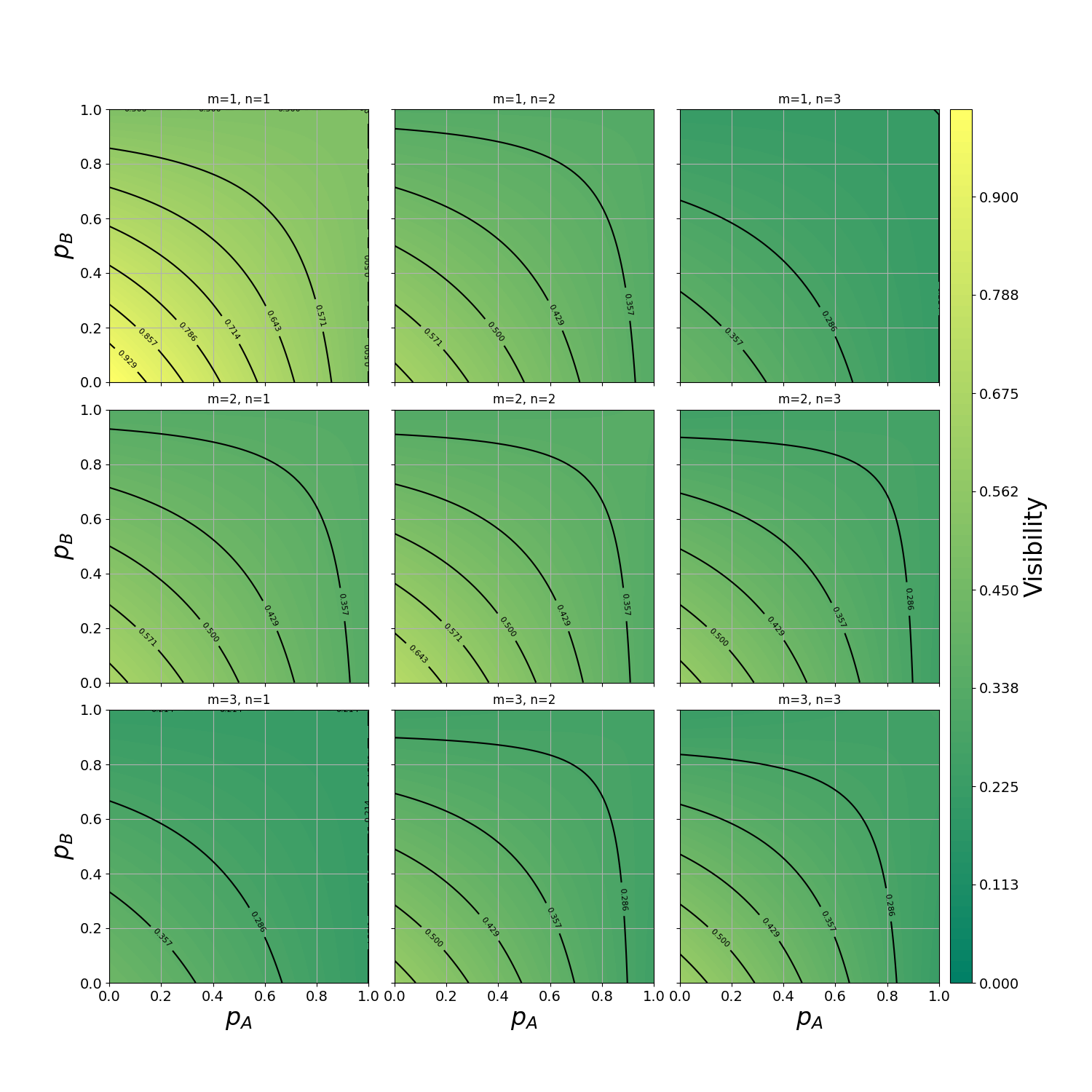}
\caption{Contour plot showing simulated HOM visibility as a function of $p_A$ and $p_B$, the depolarization probabilities, for different photon number inputs $m$ and $n$ to the channel. Depolarization leads to loss of coherence in the polarization degree of freedom, reducing HOM visibility.}
\label{fig:depolarizing_contour}
\end{figure}

Secondly, we investigate the impact of varying the depolarizing probability \(p\) on the visibility of HOM. Shown in Figure \ref{fig:depolarizing_single_photons}, when single photons, indistinguishable in all degrees of freedom beyond polarization, are input into the beam splitter for various strengths of depolarization noise, the coincidence probability is drawn to a fixed value as the strength of the depolarization increases. In addition, Figure \ref{fig:depolarizing_contour} shows how depolarization noise shapes the ability to achieve high visibility interference. Depolarization reduces the coherence of the input states which inhibits high indistinguishability and will consequently affect quantum information protocols.  Ultimately, the increased probability of depolarization reduces visibility, underscoring the need for precise polarization management in quantum optical systems.


\begin{figure}[ht!]
\centering
\includegraphics[width=\linewidth]{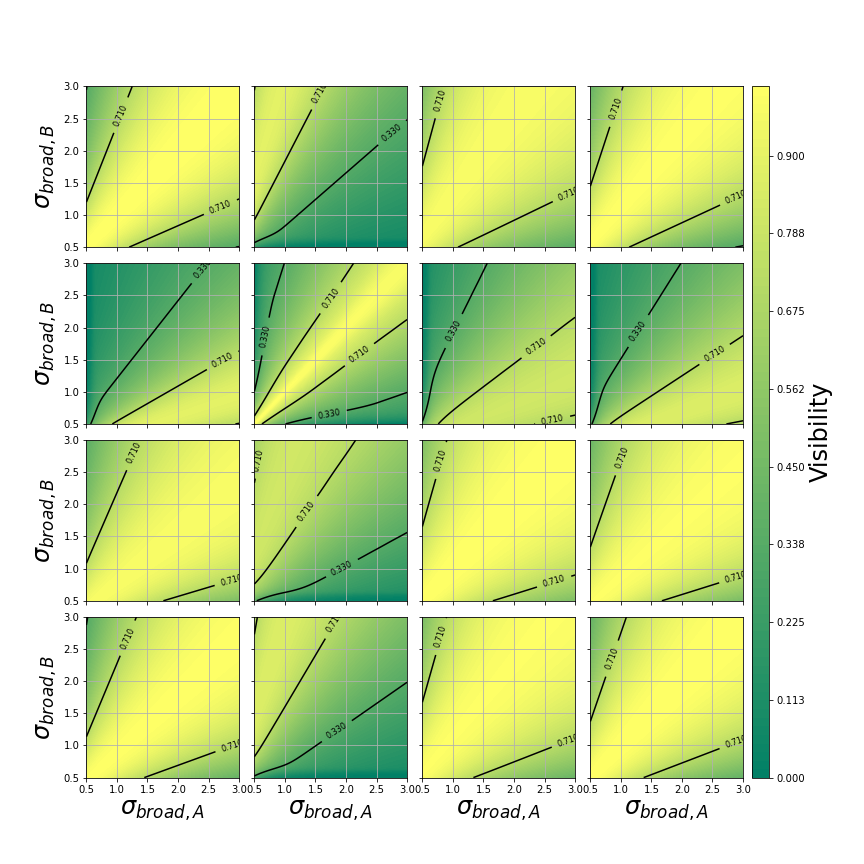}
\caption{Contour plot showing simulated HOM visibility as a function of $\Delta \lambda_A$ and $\Delta \lambda_B$, the spectral broadening strength, for single photons of the considered spectral shapes. Here spectral broadening increased the region of high visibility in the frequency domain, but due to the Fourier relationship between frequency and time, the temporal region of high visibility is restricted.}
\label{fig:spectral_broadening_contour}
\end{figure}

Thirdly, we inspect how spectral broadening alters the HOM visibility for single photons sourced from both parties. The spectral shape is varied for each input photon, but the input spectral bandwidths and central frequencies are originally matched before the source photons are passed through independent spectral broadening channels. The spectral broadening factor is varied from $0.5$ to $3.0$ for each channel. As evidenced in Figure \ref{fig:spectral_broadening_contour}, careful consideration of the effects of the channels on the spectral properties of photons is required to ensure users are operating in a regime in which high-visibility interference is achievable, as spectral broadening alters interference visibility by modifying the spectral overlap of the sourced photons. Here, spectral broadening increased the region of high visibility in the frequency domain, but due to the Fourier relationship between frequency and time, the temporal region of high visibility is restricted.

The combined effects of amplitude damping, depolarization, and spectral broadening thus strongly affect the interference visibility in quantum networking. Amplitude damping decreases the photon number in each mode, influencing the visibility of the HOM interference. In states with a higher number of photons, amplitude damping may alleviate photon number mismatches at the input and partly counterbalance some detrimental effects, thus enhancing the visibility under certain conditions. Depolarization decreases coherence between the sourced photons and reduces visibility of interference. High interference visibility relies on precise polarization control in order to limit depolarization. Finally, spectral broadening modifies the spectral distribution of the photons, which may alter the spectral overlap of the input photons, hence decreasing interference visibility. The spectral shape and bandwidth variations immediately raise concern about the quality of HOM interference and underline the management of spectral properties as an important condition for high visibility. The successive action of these quantum channels underlines the mitigation of photon loss, depolarization, and spectral mismatches as an important task for optimizing interference-based quantum information protocols.

\section{Coherent States}\label{sec:coherent}
\begin{figure}[ht!]
    \centering
    \includegraphics[scale=0.225]{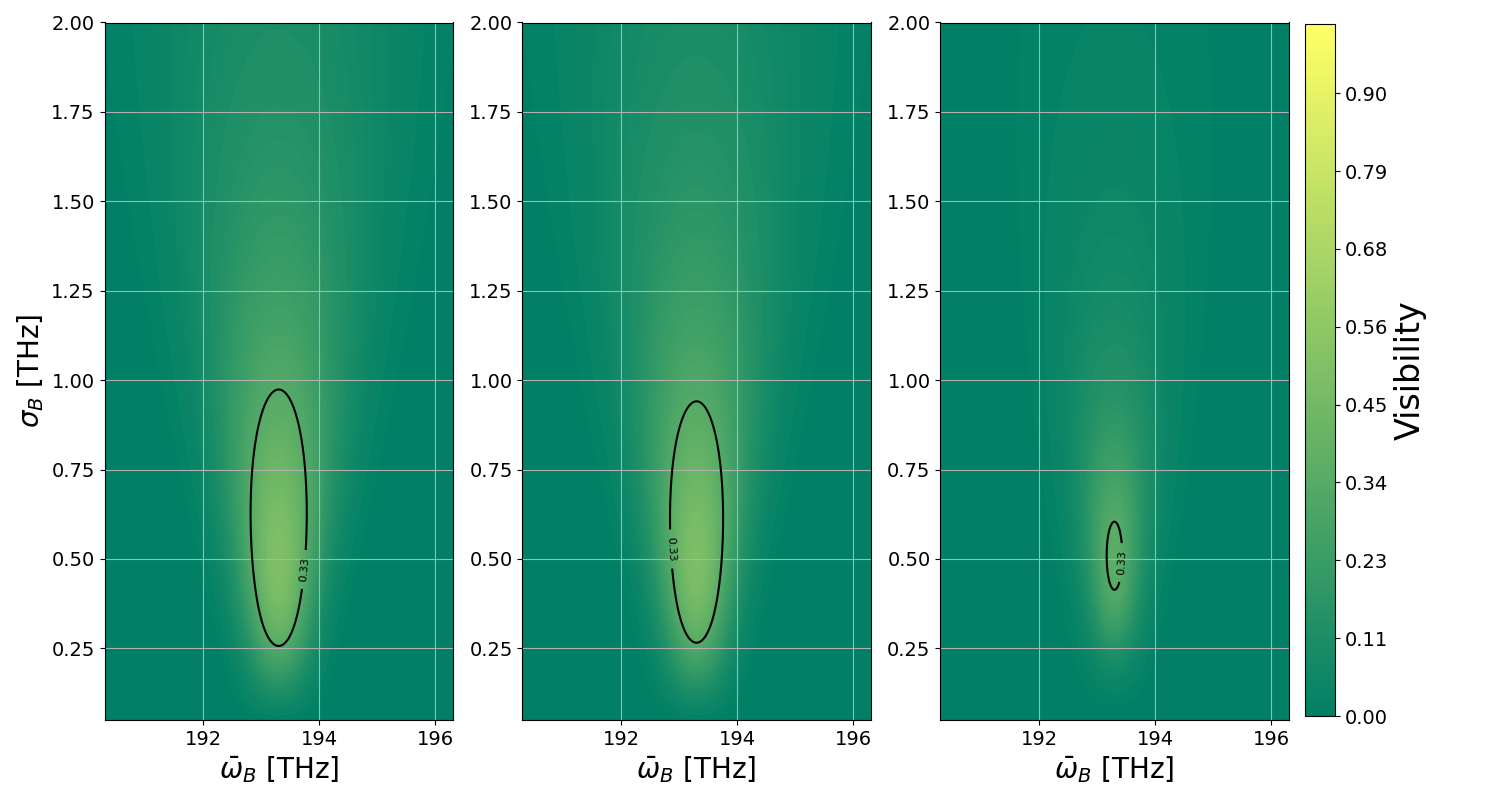}
    \caption{The contour plot depicts the simulated visibility of the HOM interference between coherent pulses. The visibility is a function of the spectral properties of input pulse B with a fixed pulse in A ($\bar{\omega}$=193.55 THz, $\sigma=0.5$ THz, $\tau=0$,  and $\cos\Phi=1$). Here from left to right $\mu_A = \mu_B = 0.1,\ 1,$ and $5$.}
    \label{fig:coh_source_vis2}
\end{figure}

\begin{figure}[ht!]
    \centering
    \includegraphics[scale=0.425]{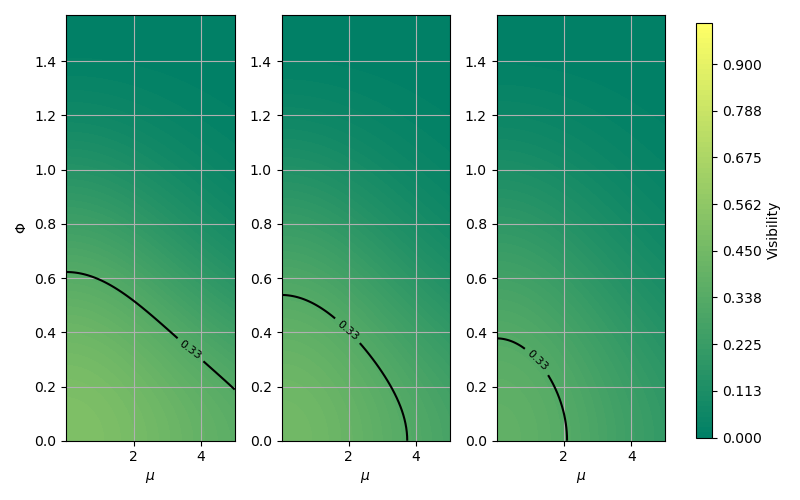}
    \caption{The contour plot depicts the simulated visibility of the HOM interference as a function of the average photon number of the sources and the polarization mismatch of photons interacting at a 50/50 beam splitter. On the left, two spectrally matched sources are used while in the middle and on the right, the coherent photons input in mode $B$ have a spectral width $1.4$ and 1.7 times that of photons in input mode $A$ respectively. As the average photon number of the sources increases, the visibility of the interference is reduced. In addition, mismatch in the polarization of photons will also reduce the HOM visibility.}
    \label{fig:coh_source_vis2a}
\end{figure}

Often times, coherent photons attenuated to the single-photon level are sourced in quantum communication protocols because they closely approximate single-photon sources while being easier to produce and manipulate. A coherent state is given by
\begin{equation}
\ket{\alpha} = \text{e}^{-\frac{|\alpha|^2}{2}} \sum_{n=0}^\infty \frac{\alpha^n}{\sqrt{n!}} \ket{n}
\end{equation}
where $\alpha = |\alpha|\text{e}^{i\theta}$, with $|\alpha|$ and $\theta$ being the amplitude and phase of the coherent state, respectively. The photon number follows a Poisson distribution with mean photon number $\mu = \langle n \rangle = |\alpha|^2$, 
\begin{equation}
    p_n (\mu ) = |\langle n|\alpha \rangle|^2 = e^{-\mu} \frac{\mu^{n}}{n!}.
\end{equation}
Most QKD schemes implement phase-randomized photon pulses for security purposes, given by the density matrix
\begin{equation}
    \rho = \frac{1}{2\pi} \int_0^{2\pi} \ket{\alpha} \bra{\alpha} d\theta = \sum_{n=0}^\infty p_n(\mu) \ket{n} \bra{n}
\end{equation}
which is a statistical mixture of photon-number eigenstates. Thus, our source can be treated as emitting photons according to a Poisson distribution $p_n (\mu)$. In the HOM setup, we have channels $A$ and $B$, each with a coherent source. The probability of emitting $m$ photons at $A$ and $n$ photons at $B$ is given by $p_m (\mu_A )$ and $p_n (\mu_B)$, respectively. So, the probability of detecting a coincidence in this instance is $p_m ({\mu_A} ) p_n ({\mu_B} ) P_{m,n}^{\text{Co}}$, where $P_{m,n}^{\text{Co}}$ is given by Eq.\ \eqref{eq:9}. The total probability of coincidence detection is obtained by summing over all the photon numbers $m,n$:
\begin{equation}\label{eq:pmu}
    P_\text{total}^{\text{Co}} = \sum_{m,n} p_m ({\mu_A} ) p_n ({\mu_B} ) P_{m,n}^{\text{Co}}
\end{equation}
For a general $T/R$ beam splitter and detector efficiency functions \eqref{eq:def}, the coincidence probability between two coherent sources producing photons with average photon numbers $\mu_A$ and $\mu_B$, respectively, is found to be
\begin{widetext}
\begin{equation}\label{eq:B5}
\begin{split}
P_{\text{total}}^{\text{Co}}     &= 1 - e^{-\mu_A \eta_A' - \mu_B \eta_B'} - e^{-\mu_A \eta_A - \mu_B \eta_B} + e^{\mu_A (\eta_A \eta_A' - \eta_A - \eta_A') + \mu_B(\eta_B \eta_B' - \eta_B - \eta_B')} \\
         &- e^{-\mu_A-\mu_B} (e^{\mu_A R + \mu_B T}+e^{\mu_A T + \mu_B R})\text{I}_0(2\sqrt{\mu_A \mu_B R T} \cos\Phi \cos\Theta) \\
         &+ e^{-\mu_A-\mu_B + \mathcal{A} + \mathcal{B}} \text{I}_0(2\sqrt{\mathcal{A}\mathcal{B}} \cos\Phi \cos\Theta) + e^{-\mu_A-\mu_B + \mathcal{C} + \mathcal{D}} \text{I}_0(2\sqrt{\mathcal{C}\mathcal{D}} \cos\Phi \cos\Theta)
    \end{split}
\end{equation}
\end{widetext}
where $I_0$ is a modified Bessel function of the first kind, $\Phi$ denotes the polarization mismatch between pulses, and $\Theta$ the mismatch in the spectral-temporal profiles of pulses $A$ and $B$, and we defined four coefficients to simplify the expression,
\begin{align}
    \mathcal{A} &= \mu_A R (1-\eta_A') \ , \
        \mathcal{B} = \mu_B T (1-\eta_B') \, \nonumber\\
        \mathcal{C} &= \mu_A T (1-\eta_A) \, \
        \mathcal{D} = \mu_B R (1-\eta_B)\ .
\end{align}
Assuming a 50/50 beam splitter, ideal detectors, and that both sources have the same average photon number, $\mu_A = \mu_B = \mu$, the coincidence probability is given by (see Appendix \ref{app:1} for details)
\begin{equation} \label{eq:coh_coin}
    P_\text{total}^{\text{Co}} = 1 + e^{-2\mu}-2e^{-\mu} I_0 \left( \mu \cos\Phi
    \cos\Theta \right).
\end{equation} 
The HOM visibility for spectrally-matched coherent sources ($\Theta =0$), following from Eqs.\ \eqref{eq:visibility} and \eqref{eq:coh_coin}, is
\begin{equation}
    \mathcal{V}(\mu, \Phi) = \frac{{I}_0 (\mu \cos \Phi) - 1}{2 \sinh^2\frac{\mu}{2}}
\end{equation}
and is plotted in Figure \ref{fig:coh_source_vis2a}, where the leftmost figure depicts the scenario of perfect spectral overlap and the images to its right introduce mismatch in the spectral bandwidth of the pulses for Gaussian shaped spectral envelopes. The ratio $\frac{\sigma_A}{\sigma_B} $ is $0.714$ (middle) and $0.588$ (right). The maximum visibility of the interference using coherent sources is $0.5$. Increasing the average photon number of the sources or the polarization mismatch degrades the visibility, as shown in Figure \ref{fig:coh_source_vis2}. 

\begin{figure}
    \centering
    \includegraphics[width=0.97\linewidth]{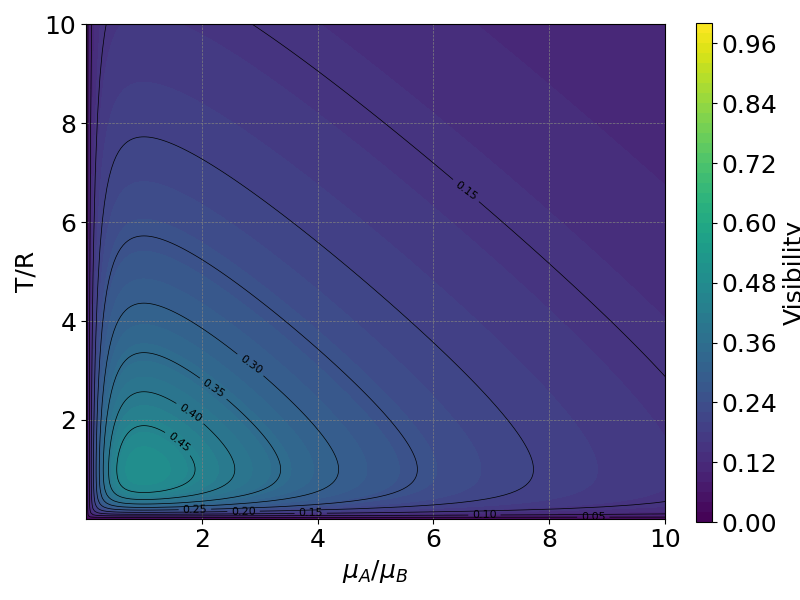}
    \caption{The simulated HOM visibility for coherent photons as a function of the input photon number ratio ($\mu_A$/$\mu_B$) vs the beam splitter transitivity, reflectivity ratio ($T$/$R$) for ideal detector efficiencies. Here the maximum visibility occurs when both ratios are equal to $1$. }
    \label{fig:enter-label}
\end{figure}
\begin{figure}
    \centering
    \includegraphics[width=0.97\linewidth]{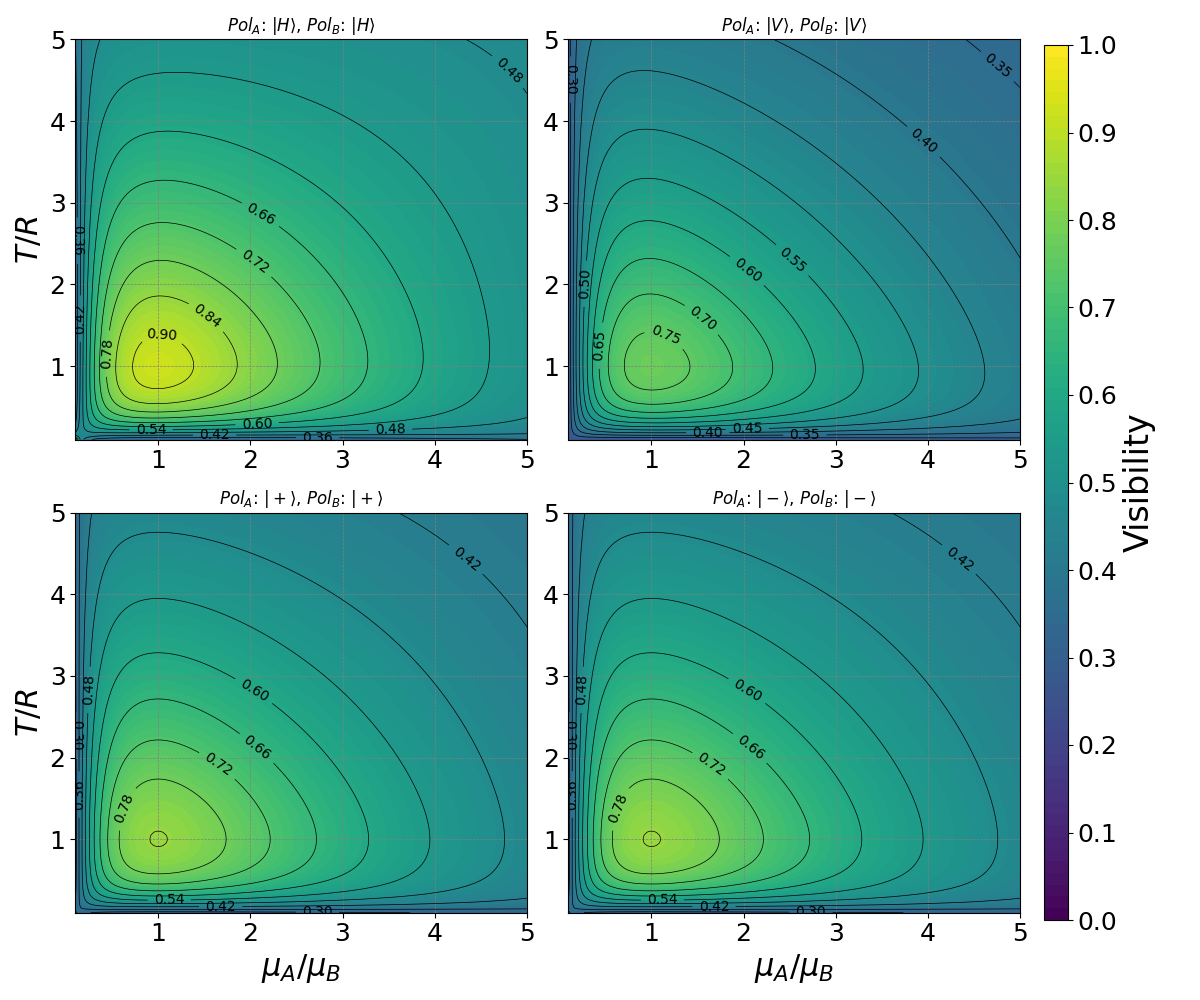}
    \caption{The simulated coherent HOM visibility for different polarization preparations as a function of the input photon number ratio ($\mu_A$/$\mu_B$) vs the beam splitter transitivity, reflectivity ratio ($T$/$R$) for reduced efficiency polarization dependent detectors ($\eta_{H,A} = 0.8,\ \eta_{V,A} = 0.83,\ \eta_{H,B} = 0.78,\ \text{and } \eta_{V,B} = 0.85$).}
    \label{fig:enter-label}
\end{figure}

\section{Applications}
\label{sec:5}
Numerous applications in quantum networks and computing rely on successful observation of photon interference, a more comprehensive model of the interference of photons that attends to spectral-temporal properties, polarization, detector efficiency, channel effects, and photon number can inform numerous quantum applications in which HOM interference plays a role. For entanglement distribution, the model allows further management of indistinguishability among photons across modes, providing higher fidelity entangled states. In MDI-QKD, improvements in interference visibility reduce the QBER and consequently enhance the secure key generation rate. For quantum optical classifiers, their accuracy is increased by minimizing mode mismatch thanks to the better modeling of interference effects over multiple degrees of freedom. More completely, the model allows for more robust protocols to be designed in photonic quantum computing for operations with photons, enhancing gate fidelities and allowing for improved scaling of quantum photonic networks.

 \subsection{Optical Bell State Measurement and Distributing Entanglement}
The HOM effect is foundational to OBSM, a key process in quantum networking protocols such as entanglement swapping. The success of OBSM critically depends on the indistinguishability of photons across all degrees of freedom, particularly polarization and spectral modes. Our multi-mode, multi-photon HOM model reveals that, when combined, mismatches across degrees of freedom can significantly degrade interference visibility, thereby reducing the probability of successful Bell state identification.

Here, we consider two independent photon pairs, systems $AB$ and $CD$, each initially prepared in the multi-mode $\ket{\Psi^-}$-like Bell state. These states exhibit entanglement in both polarization and frequency. By performing a polarization-resolving BSM on photons $B$ and $C$, entanglement is "swapped" to photons $A$ and $D$, producing a hybrid entangled state across polarization and frequency. In general, it is not possible to exceed the total success probability of 50\% with linear optical elements \cite{doi:10.1126/sciadv.adf4080}. Below, we define the initial states and lay the groundwork for the swapping protocol.

Firstly, we will define a set of single-frequency-pair Bell state operators to simplify the following expressions, over $AB$ these are:
\begin{equation}
    \begin{split}
        &\hat{\Psi}^{\pm\ \dagger}_{AB}(\omega_A, \omega_B) = \frac{1}{\sqrt{2}}\Big(\hat{a}^\dagger_H(\omega_A) \hat{b}^\dagger_V(\omega_B) \pm \hat{a}^\dagger_V(\omega_A) \hat{b}^\dagger_H(\omega_B)\Big) \\
        &\hat{\Phi}^{\pm\ \dagger}_{AB} (\omega_A, \omega_B) = \frac{1}{\sqrt{2}}\Big(\hat{a}^\dagger_H(\omega_A) \hat{b}^\dagger_H(\omega_B) \pm \hat{a}^\dagger_V(\omega_A) \hat{b}^\dagger_V(\omega_B)\Big).
    \end{split}
\end{equation}
The Bell operators $\hat{\Psi}^{\pm\ \dagger}_{AB}(\omega_A, \omega_B)$ and $\hat{\Phi}^{\pm\ \dagger}_{AB} (\omega_A, \omega_B)$ create Bell-type entangled states where the two photons occupy distinct frequency modes $\omega_A$ and $\omega_B$.  Similar operators can be defied for the $CD$ system. 

Spontaneous Parametric Down-Conversion (SPDC) in a nonlinear crystal is a fundamental method for generating entangled photon pairs. In type-II SPDC, the signal and idler photons emerge with orthogonal polarizations, facilitating the creation of polarization-entangled states. The joint spectral properties of these photons, encapsulated in the Joint Spectral Amplitude (JSA), critically influence both the degree of spectral entanglement and the fidelity of quantum operations like entanglement swapping. The spectral properties of the sourced photons are determined by the phase-matching conditions of the nonlinear process and energy-conservation. Thus, type-II SPDC gives rise to a distribution of single-frequency-pair Bell states with $\Psi^-$-like entanglement, parameterized by the JSA $f(\omega, \omega')$. If we consider independent Type-II SPDC sources with JSAs $f_{AB}(\omega_A, \omega_B)$ and $f_{CD}(\omega_C, \omega_D)$ respectively, the initial states of the photon pairs $AB$ and $CD$ when written in terms of the defined Bell operators:
\begin{equation}
    \begin{split}
        &\ket{\psi_{AB}}=\int \int \text{d}\omega_A\ \text{d}\omega_B\ f_{AB}(\omega_A, \omega_B) \hat{\Psi}^{-\ \dagger}_{AB}(\omega_A, \omega_B) \ket{0}_{AB} \\
        & \ket{\psi_{CD}}=\int \int \text{d}\omega_C\ \text{d}\omega_D\ f_{CD}(\omega_C, \omega_D) \hat{\Psi}^{-\ \dagger}_{CD}(\omega_C, \omega_D) \ket{0}_{CD}
    \end{split}
\end{equation}
or expanded fully in terms of spatial, polarization and frequency modes,
\begin{align}
    &\ket{\psi_{AB}} = \frac{1}{\sqrt{2}} \int \int \text{d}\omega_A\ \text{d}\omega_B\ f_{AB}(\omega_A, \omega_B) \\
    &\quad\quad\quad\quad\quad\big{(}\hat{a}^\dagger_H(\omega_A)\hat{b}^\dagger_V(\omega_B) - \hat{a}^\dagger_V(\omega_A) \hat{b}^\dagger_H(\omega_B) \big{)} \ket{0_{AB}} \\
    &\ket{\psi_{CD}} = \frac{1}{\sqrt{2}} \int \int \text{d}\omega_C\ \text{d}\omega_D\ f_{CD}(\omega_C, \omega_D) \\
    &\quad\quad\quad\quad\quad\big{(}\hat{c}^\dagger_H(\omega_C)\hat{d}^\dagger_V(\omega_D) - \hat{c}^\dagger_V(\omega_C) \hat{d}^\dagger_H(\omega_D) \big{)} \ket{0_{CD}}.
\end{align}

The JSA factorizes into a pump envelope function (PEF), $\mathcal{E}(\omega_s, \omega_i)$, and a phase-matching function (PMF), $\phi(\omega_s, \omega_i)$. The PEF, determined by the pump’s temporal and spectral profile, sets the overall bandwidth and frequency correlations. The PMF, governed by the crystal’s birefringence, dispersion, and poling structure, imposes constraints that must be satisfied for efficient frequency conversion. Under Gaussian assumptions, The PEF is centered at frequency $\omega_p$ with a bandwidth $\sigma_p$, its spectral form is 
\begin{equation}
\mathcal{E}(\omega_s, \omega_i) = \exp\left(-\frac{(\omega_s + \omega_i - \omega_p)^2}{2 \sigma_p^2}\right),
\end{equation}
where $\omega_s$ and $\omega_i$ are the signal and idler frequencies, respectively. The central condition $\omega_s + \omega_i = \omega_p$ reflects energy conservation. The bandwidth $\sigma_p$ governs the spectral correlation between the signal and idler photons, with broader pump bandwidths leading to stronger spectral entanglement. For a Gaussian approximation of the PMF, the functional form is
\begin{equation}
\phi(\omega_s, \omega_i) = \exp\left(-\frac{\Delta k^2}{2 \sigma_{\text{pm}}^2}\right),
\end{equation}
where $\Delta k$ is the phase mismatch, and $\sigma_{\text{pm}}$ describes the width of the phase-matching window. The phase mismatch $\Delta k$ is typically a function of the signal and idler frequencies and depends on the slopes of the crystal's dispersion curves.

Spectral entanglement arises when the JSA cannot be factorized into separate signal and idler components. Such correlations influence the performance of entanglement swapping which relies on the coherent interference of photons from independent entangled photon pair source such as SPDC sources. Mismatches in their JSAs due to differing pump bandwidths or crystal conditions introduce spectral distinguishability, reducing the fidelity of the swapped entangled states. 

The joint state of the system is then,
\begin{equation}
    \ket{\psi} = \ket{\psi_{AB}} \otimes \ket{\psi_{CD}}
\end{equation}
Systems $AB$ and $CD$ prepare their photon pairs independently, and this independence can introduce misalignment between the optical axes of each subsystem. The polarization mismatch can be introduced by applying a rotation to the polarization of one of the photons, say photon \(C\). Let us assume that photon \(C\) undergoes a polarization rotation by an angle $(\Phi)$. The creation operators for photon \(C\) are then transformed as follows:
\begin{equation}
    \begin{pmatrix}
        \hat{c}_H^\dagger(\omega) \\
        \hat{c}_V^\dagger(\omega)
    \end{pmatrix} 
    \rightarrow 
    \begin{pmatrix}
        \cos(\Phi) & -\sin(\Phi) \\
        \sin(\Phi) & \cos(\Phi)
    \end{pmatrix}
    \begin{pmatrix}
        \hat{c}_H^\dagger(\omega) \\
        \hat{c}_V^\dagger(\omega)
    \end{pmatrix}.
\end{equation}
Applying this transformation to our initial state gives rise to the transformed operators $\underline{\hat{c}}_i^\dagger$ where $i \in \{H,V\}$. 
\begin{equation}
\begin{split}
    \ket{\psi} = \frac{1}{2} \int \cdots \int &\text{d}\omega_A\ \text{d}\omega_B\ \text{d}\omega_C\ \text{d}\omega_D\ f_{AB}(\omega_A, \omega_B) f_{CD}(\omega_C, \omega_D) \\
    &\times \big{[}\hat{a}^\dagger_H(\omega_A) \hat{b}^\dagger_V(\omega_B) - \hat{a}^\dagger_V(\omega_A) \hat{b}^\dagger_H(\omega_B)\big{]} \\
    &\otimes \big{[}\underline{\hat{c}}^\dagger_H(\omega_C) \hat{d}^\dagger_V(\omega_D) - \underline{\hat{c}}^\dagger_V(\omega_C) \hat{d}^\dagger_H(\omega_D)\big{]}\ket{0}_{ABCD}
\end{split}
\end{equation} 
To perform an OBSM in the linear optics regime a rotation is applied to the input photons $B$ and $C$ via a beam splitter. It is crucial for the photons to enter each limb of the beam splitter coincidentally, and in order to achieve quantum interference they must have a high degree of indistinguishability.  Thus, photons \(B\) and \(C\) are sent to a 50:50 beam splitter which transforms, for $i\in\{ V,H \}$, the input modes according to the relations:
\begin{equation}
\begin{split}
    &\hat{b}_{i} (\omega) \rightarrow \frac{1}{\sqrt{2}} \left( \hat{b}_{i} (\omega) + \hat{c}_{i} (\omega) \right), \\
    &\hat{c}_{i} (\omega) \rightarrow \frac{1}{\sqrt{2}} \left( \hat{b}_{i} (\omega) - \hat{c}_{i} (\omega) \right).
    \end{split}
\end{equation}
Applying these transformations to the state \(\ket{\psi}\), we obtain $\ket{\Psi_{BS}}$, the transformed state involving modes \(A, B, C,\) and \(D\), incorporating the spectral and polarization properties of the photons. Following the initial beam splitter,  a polarizing beam splitter (PBS) is introduced at each output port of the beam splitter and four detectors are located across the output ports of the PBSs. To see the combined effect of spectral and polarization mismatch on the resultant fidelity of the entangled state between photons \(A\) and \(D\), we consider the following four measurement outcomes:
\begin{equation}\label{eq:M0}
\begin{split}
    M_0 &= \int d\omega d\omega' \hat{b}^\dagger_{H}(\omega)  \hat{c}^\dagger_{V}(\omega') |0\rangle_{BC} \langle 0|  \hat{c}_{V}(\omega')  \hat{b}_{H}(\omega) \\
    M_1 &= \int d\omega d\omega' \hat{c}^\dagger_{H}(\omega)  \hat{b}^\dagger_{V}(\omega') |0\rangle_{BC} \langle 0|  \hat{b}_{V}(\omega')  \hat{c}_{H}(\omega) \\
    M_2 &= \int d\omega d\omega' \hat{b}^\dagger_{H}(\omega)  \hat{b}^\dagger_{V}(\omega') |0\rangle_{BC} \langle 0|  \hat{b}_{V}(\omega')  \hat{b}_{H}(\omega) \\
    M_3 &= \int d\omega d\omega' \hat{c}^\dagger_{H}(\omega)  \hat{c}^\dagger_{V}(\omega') |0\rangle_{BC} \langle 0|  \hat{c}_{V}(\omega')  \hat{c}_{H}(\omega) \\
\end{split}
\end{equation}

The outcomes $M_i$ occur with equal probabilities, $p_0 = p_1=p_2=p_3$, where
\begin{equation}
    p_i = \| \bra{\Psi_{\text{BS}}} M_i \ket{\Psi_{\text{BS}}} \|^2 = \frac{1}{8}
\end{equation}
Focusing on the measurement $M_0$, we define a new set of single-frequency-pair Bell state operators now over $AD$ to simplify the following expressions:
\begin{equation}
    \begin{split}
        &\hat{\Psi}^{\pm\ \dagger}_{AD}(\omega_A, \omega_D) = \frac{1}{\sqrt{2}}\Big(\hat{a}^\dagger_H(\omega_A) \hat{d}^\dagger_V(\omega_D) \pm \hat{a}^\dagger_V(\omega_A) \hat{d}^\dagger_H(\omega_D)\Big) \\
        &\hat{\Phi}^{\pm\ \dagger}_{AD} (\omega_A, \omega_D) = \frac{1}{\sqrt{2}}\Big(\hat{a}^\dagger_H(\omega_A) \hat{d}^\dagger_H(\omega_D) \pm \hat{a}^\dagger_V(\omega_A) \hat{d}^\dagger_V(\omega_D)\Big).
    \end{split}
\end{equation}
The Bell operators $\hat{\Psi}^{\pm\ \dagger}_{AD}(\omega_A, \omega_D)$ and $\hat{\Phi}^{\pm\ \dagger}_{AD} (\omega_A, \omega_D)$ create Bell-type entangled states where the two photons occupy distinct frequency modes $\omega_A$ and $\omega_D$. Similarly, we define 
\begin{equation}
    \begin{split}
        \hat{\chi}_{BC}^{- \dagger}(\omega_B, \omega_C) &= \frac{1}{\sqrt{2}} \left( \hat{b}^\dagger_H (\omega_C) \, \hat{c}^\dagger_V (\omega_B) - \hat{b}^\dagger_H (\omega_B) \, \hat{c}^\dagger_V (\omega_C) \right) \\
        \hat{\chi}_{BC}^{+ \dagger}(\omega_B, \omega_C) &= \frac{1}{\sqrt{2}} \left( \hat{b}^\dagger_H (\omega_C) \, \hat{c}^\dagger_V (\omega_B) + \hat{b}^\dagger_H (\omega_B) \, \hat{c}^\dagger_V (\omega_C) \right)
    \end{split}
\end{equation}
where $-$ and $+$ represent the symmetric and anti-symmetric combinations of the particular mode operators respectively. Then we may denote the states:
\begin{equation}
    \begin{split}
        & \ket{\chi_{BC}^-(\omega_B, \omega_C)} = \hat{\chi}_{-}^\dagger(\omega_B, \omega_C) \ket{0}_{BC} \\
        & \ket{\chi_{BC}^+(\omega_B, \omega_C)} = \hat{\chi}_{+}^\dagger(\omega_B, \omega_C) \ket{0}_{BC}
    \end{split}
\end{equation}

Expressing in terms of these operators gives
\begin{equation}
\begin{split}
    \hat{M}_0\ket{\Psi_{BS}} = &\frac{1}{2} \int \cdots \int\mathrm{d}\omega_A\ \mathrm{d}\omega_B\ \mathrm{d}\omega_C\ \mathrm{d}\omega_D\ f_{AB}(\omega_A, \omega_B) f_{CD}(\omega_C, \omega_D)   \\
    &\times \Big[ \cos\Phi \Big( \ket{\Psi^+_{AD} (\omega_A, \omega_D)} \ket{\chi^-_{BC}(\omega_B, \omega_C)} \\
    &\quad \quad + \ket{\Psi^-_{AD}  (\omega_A, \omega_D)} \ket{\chi^+_{BC}(\omega_B, \omega_C)}\Big) \\
    &\quad - \sin\Phi \Big( \ket{\Phi^+_{AD} (\omega_A, \omega_D)} \ket{\chi^+_{BC}(\omega_B, \omega_C)}  \\
    &\quad \quad + \ket{\Phi^-_{AD} (\omega_A, \omega_D)} \ket{\chi^-_{BC}(\omega_B, \omega_C)} \Big]
\end{split}
\end{equation}
The post measurement state is then,
\begin{equation}
    \begin{split}
     \ket{\psi^\text{final}} &= \frac{ M_0\ket{\Psi_{\mathrm{BS}}}}{\sqrt{\bra{\Psi_{\text{BS}}} M_i \ket{\Psi_{\text{BS}}}}}
    \end{split}
\end{equation}

The resultant state of the $AD$ system, $\rho_{AD}$, is found by tracing out the $BC$ subsystem from the final state.  For a polarization encoded scheme, we are interested in the successful swapping of entanglement across the polarization degree of freedom for $A$ and $D$. We can access the similarity of the polarization entanglement of the resultant states with that of a $\ket{\Psi^-}$ Bell state via fidelity. In order to do so, we must trace out the spectral degree of freedom for the $AD$ system. Doing so, one arrives at an expression for the fidelity
\begin{equation}
\begin{split}
     &F(\rho_{AD}, \ket{\Psi^-}\bra{\Psi^-}) = \frac{\cos^2\Phi}{2} \\
     &\Big[1 + \int\int d\omega_A\ d\omega_D\ F_{AB,CD}(\omega_A, \omega_D)\ F_{CD,AB}(\omega_D, \omega_A)\Big]
\end{split}
\end{equation}
where we define the overlap integrals
\begin{equation}
    \begin{split}
        &F_{AB,CD}(\omega_A, \omega_D) = \int d\omega_B f_{AB}(\omega_A, \omega_B) f_{CD}^*(\omega_B, \omega_D) \\
        &F_{CD,AB}(\omega_D, \omega_A) = \int d\omega_C f_{CD}(\omega_C, \omega_D) f_{AB}^*(\omega_A, \omega_C)
    \end{split}
\end{equation}
In the case of separable JSAs, the expression for fidelity simplifies to
\begin{equation} \label{eq:fidelity}
    F(\rho^{\text{separable}}_{AD}, \ket{\Psi^-}\bra{\Psi^-}) = \frac{\cos^2\Phi}{2} \Big[1 + \cos^2\Theta_{BC}\Big]
\end{equation}

\begin{figure*}[htbp]
    \centering
    \includegraphics[width=\textwidth]{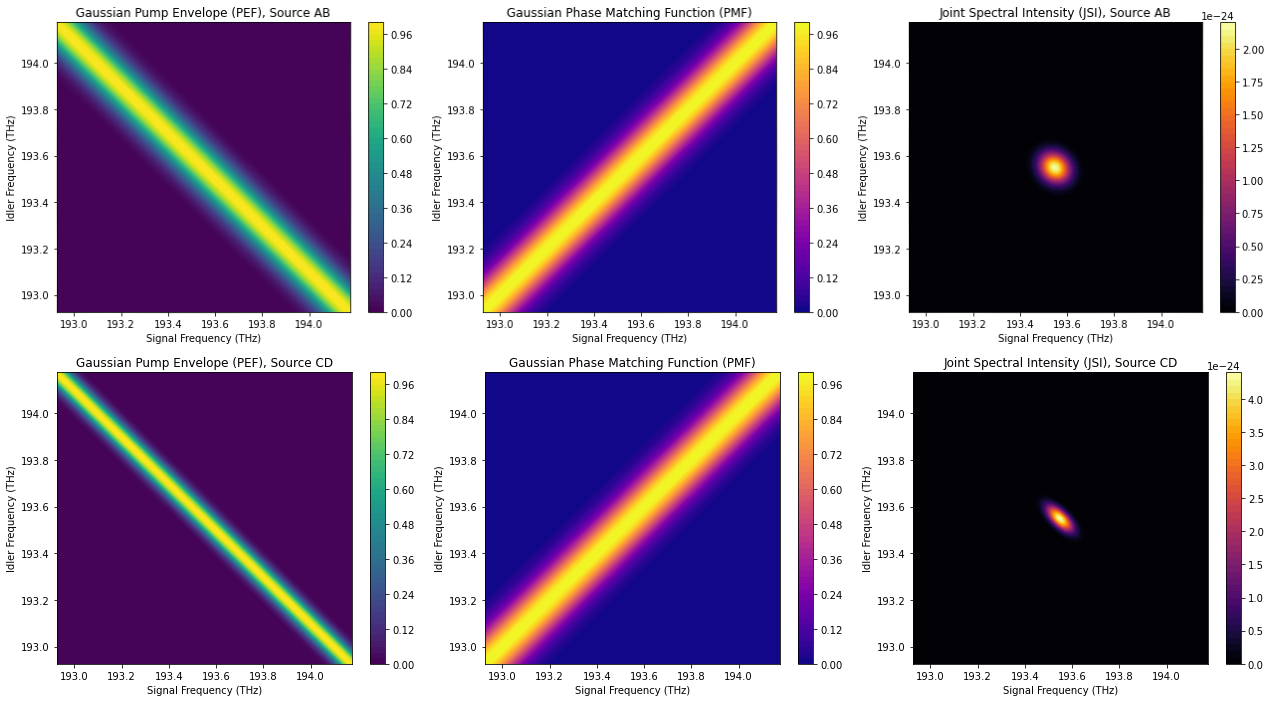} 
    \caption{The PEF, PMF, and JSI for two independent SPDC sources. Source BC has a more narrowly defined pump bandwidth which contributes to the presence of spectral entanglement of the source photons when combined with the PMF.}
    \label{fig:spdc_JSAs}
\end{figure*}

\begin{figure}
    \centering
    \includegraphics[width=1\linewidth]{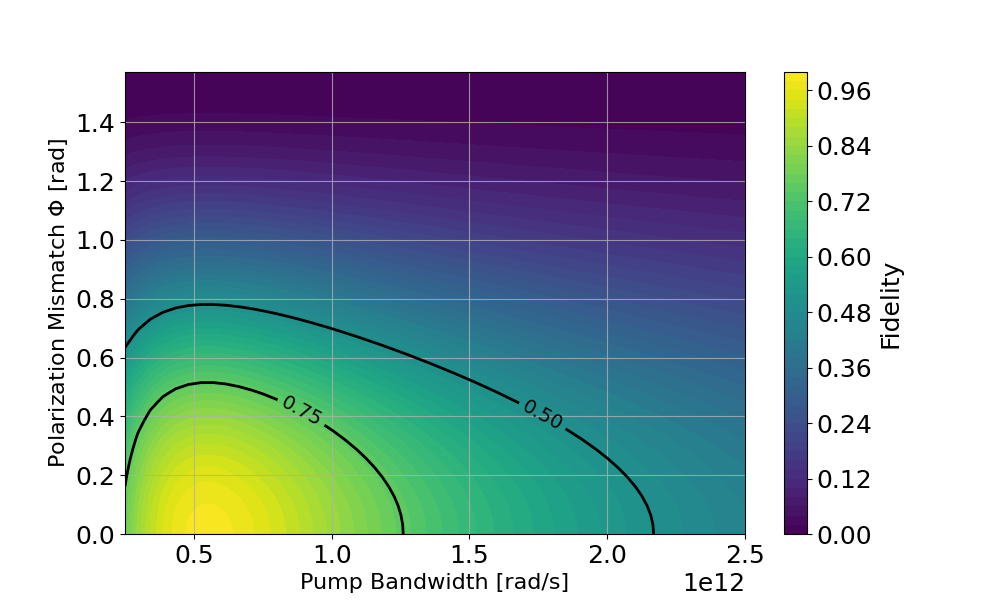}
    \caption{Post BSM fidelity of the $AD$ photons with $\ket{\Psi^-}$ as a function of SPDC pump bandwidth and polarization mismatch.}
    \label{fig:enter-label}
\end{figure}

In the case where photon pairs $AB$ and $CD$ have separable JSAs and Gaussian spectral profiles for photons \(B\) and \(C\):

\begin{equation}
    \phi_B(\omega) = \frac{1}{(\sigma_B \sqrt{\pi})^{1/2}} e^{- \frac{(\omega - \omega_{0,B})^2}{2 \sigma_B^2}},
\end{equation}

\begin{equation}
    \phi_C(\omega) = \frac{1}{(\sigma_C \sqrt{\pi})^{1/2}} e^{- \frac{(\omega - \omega_{0,C})^2}{2 \sigma_C^2}}.
\end{equation}

The overlap integral is then given by:

\begin{equation}
    \int_{-\infty}^{\infty} d\omega \, \phi_B^*(\omega) \phi_C(\omega) = \left( \frac{2 \sigma_B \sigma_C}{\sigma_B^2 + \sigma_C^2} \right)^{1/2} e^{- \frac{(\omega_{0,B} - \omega_{0,C})^2}{2 (\sigma_B^2 + \sigma_C^2)}}.
\end{equation}

\begin{figure}
    \centering
    \includegraphics[width=1\linewidth]{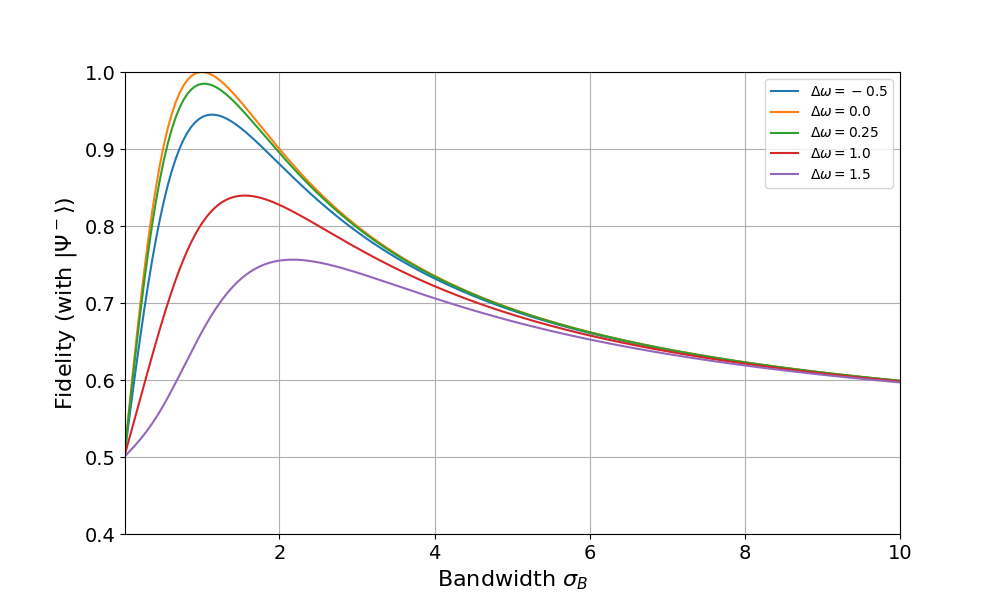}
    \caption{Post BSM fidelity with $\ket{\Psi^-}$ for photons with various spectral properties when sourcing spectrally separable photons. The bandwidth of one input photon, $\sigma_B$, is varied with respect to the other photon's bandwidth, $\sigma_C$. The different color curves indicate the difference between input photons central frequencies where, $\omega_{0,C} =1.0$ (arbitrary units) in all cases depicted.}
    \label{fig:separable-JSA-ES}
\end{figure}

This expression shows that the overlap integral—and therefore fidelity—decreases as the central frequencies \(\omega_{0,B}\) and \(\omega_{0,C}\) differ or as the bandwidths \(\sigma_B\) and \(\sigma_C\) differ. Mismatches in the central frequency $\omega_0$ and the bandwidth of sourced photons reduce the overall achievable fidelity. In the presence of central frequency mismatch, the bandwidth mismatch can be tuned to approach a point of optimal fidelity, as seen in Figure \ref{fig:separable-JSA-ES}. These techniques are crucial in the design of high-fidelity quantum networks and long-distance quantum communication protocols.

\begin{figure}
    \centering
    \includegraphics[width=1\linewidth]{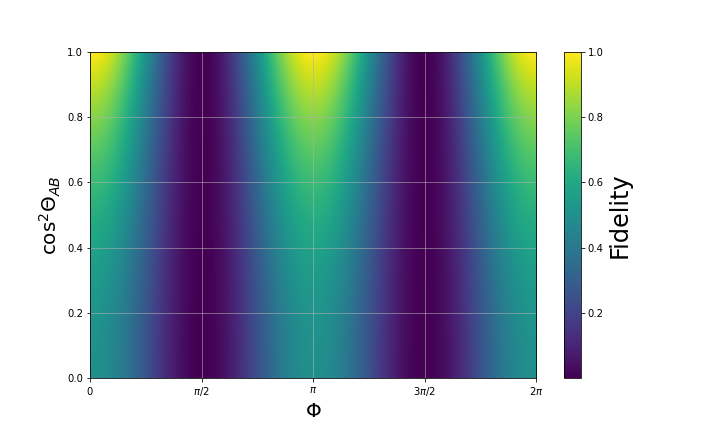}
    \caption{Simulated entanglement swapping fidelity as a function of the polarization and spectral mismatch for photons with separable JSA.}
    \label{fig:enter-label}
\end{figure}

In quantum networks with heterogeneous photon sources, polarization and spectral mismatches between different subsystems are inevitable. Our model demonstrates that managing these mismatches is critical to ensuring the success of OBSM and high-fidelity entanglement distribution. Active strategies for mode matching, including spectral filtering and polarization control, are essential to overcome these practical challenges. As quantum networks scale, robust mode-matching techniques will be crucial for maintaining high-quality entanglement distribution across network nodes.

\subsection{MDI-QKD}
MDI-QKD protocols \cite{Lo_2012, Wang2013, Woodward2021}, which are designed to eliminate vulnerabilities from detection devices, rely heavily on HOM interference to perform Bell state measurements between photons sent from two distant parties, Alice and Bob. MDI-QKD can be implemented using polarization-encoded qubits (suitable for free-space communication) \cite{Silva_MDI_pol_encode, Reaz_2024} or time-bin-encoded qubits \cite{Chan:14, Woodward2021} (suitable for fiber-integrated communication). In this context, the visibility of HOM interference directly influences the quantum bit error rate (QBER) and, subsequently, the key generation rate. Our multi-mode interference model is well-suited for analyzing how imperfections in the spectral and polarization modes of the photons degrade interference visibility, and thus the overall performance of MDI-QKD systems.

In the polarization encoding scheme, Alice and Bob send one of the states $\{ \ket{H}, \ket{V}, \frac{1}{\sqrt{2}} ( \ket{H} \pm \ket{V} ) \}$ to a central relay (Charles) where a Bell state measurement is performed. Of interest are the cases in which Charles receives one of these four states:

\begin{figure}[h]
    \centering
    \includegraphics[width=0.6\linewidth]{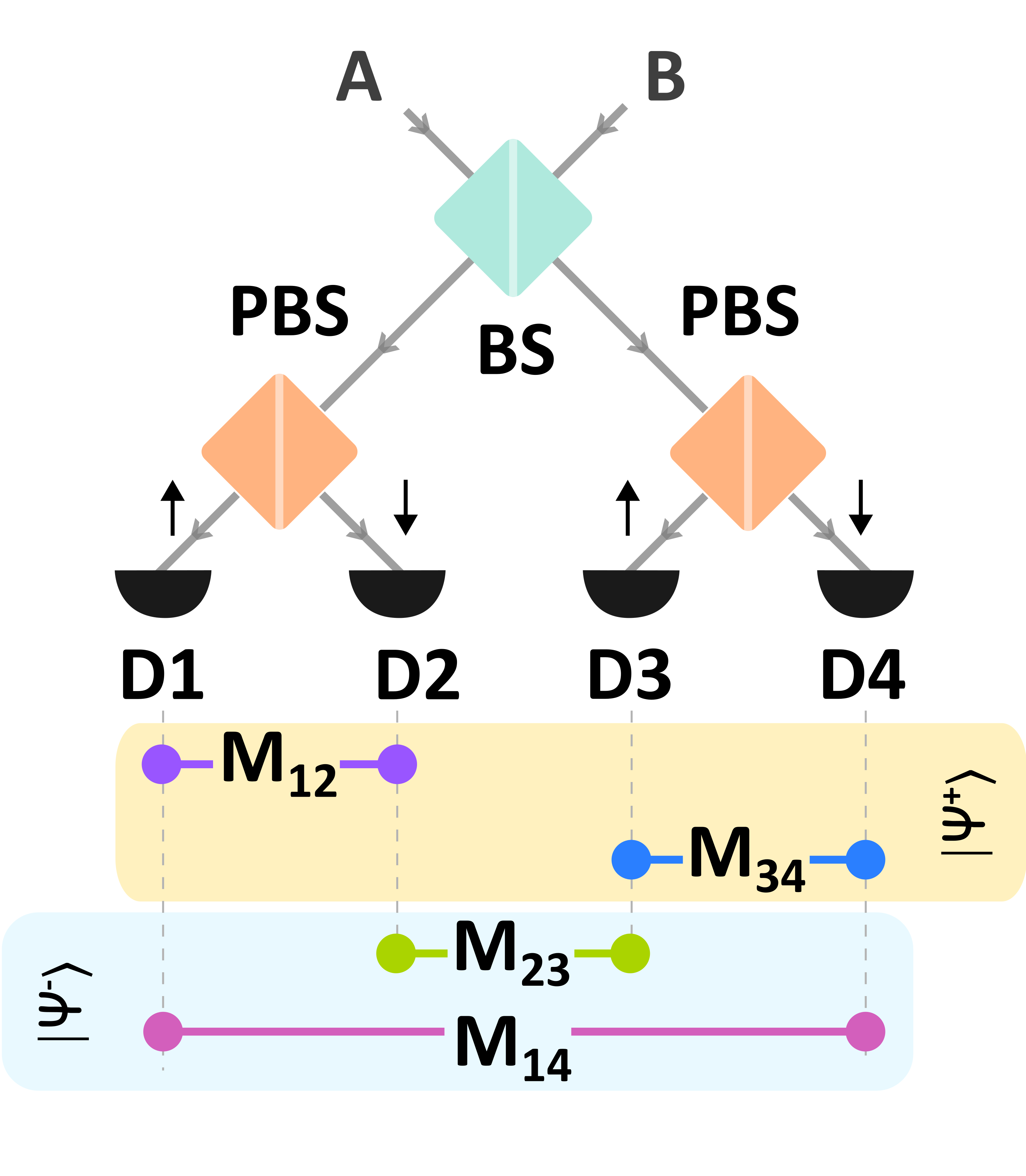}
    \caption{ Measurement setup for MDI-QKD. The polarization beam splitters (PBS) are calibrated for rectilinear basis, hence the measurements of the non-orthogonal states in either basis are discarded. Only the orthogonal states in diagonal basis can provide valid secret key through the measurements $M_3$ and $M_4$. The success probability/secure key rate depends on the associated HOM visibilities.}
    \label{fig:MDI-QKD}
\end{figure}

\begin{align}
    &\hat{A}_H^\dagger [\phi_A] \hat{B}_V^\dagger [\phi_B]  \ket{0} \ , \ \ \hat{A}_V^\dagger [\phi_A] \hat{B}_H^\dagger [\phi_B]  \ket{0} \ , \nonumber\\ &\frac{1}{2} \left( \hat{A}_H^\dagger [\phi_A] \pm \hat{A}_V^\dagger [\phi_A]\right) \left( \hat{B}_H^\dagger [\phi_B] \mp \hat{B}_V^\dagger [\phi_B]\right) \ket{0}
\end{align}
Charles sends the two photons to a 50:50 beam splitter which

transforms the input modes according to the relations:
\begin{align}\label{eq:31bs}
    \hat{a}_{i} (\omega) &\rightarrow \frac{1}{\sqrt{2}} \left( \hat{a}_{i} (\omega) + \hat{b}_{i} (\omega) \right), \nonumber\\
    \hat{b}_{i} (\omega) &\rightarrow \frac{1}{\sqrt{2}} \left( \hat{a}_{i} (\omega) - \hat{b}_{i} (\omega) \right), 
\end{align}
$i\in\{ V,H \}$, and then uses four detectors to obtain these four measurement outcomes:
\begin{equation}\label{eq:M1}
\begin{split}
    M_{34} &= \int d\omega d\omega' \hat{a}^\dagger_{H}(\omega)  \hat{a}^\dagger_{V}(\omega') |0\rangle \langle 0|  \hat{a}_{V}(\omega')  \hat{a}_{H}(\omega) \\
    M_{12} &= \int d\omega d\omega' \hat{b}^\dagger_{H}(\omega)  \hat{b}^\dagger_{V}(\omega') |0\rangle \langle 0|  \hat{b}_{V}(\omega')  \hat{b}_{H}(\omega) \\
    M_{23} &= \int d\omega d\omega' \hat{a}^\dagger_{H}(\omega)  \hat{b}^\dagger_{V}(\omega') |0\rangle \langle 0|  \hat{b}_{V}(\omega')  \hat{a}_{H}(\omega) \\
    M_{14} &= \int d\omega d\omega' \hat{b}^\dagger_{H}(\omega)  \hat{a}^\dagger_{V}(\omega') |0\rangle \langle 0|  \hat{a}_{V}(\omega')  \hat{b}_{H}(\omega) 
\end{split}
\end{equation}
$M_{12, 34}$ ($M_{23, 14}$) effectively project onto the Bell state $\ket{\Psi^+}_{AB}$ ($\ket{\Psi^-}_{AB}$). For the rectilinear basis, all outcomes are equally probable, with probability $\frac{1}{4}$ each. For the diagonal basis, only outcomes $M_{23, 14}$ are conclusive. Considering the fidelity factor for separable spectra from Eq.\ \eqref{eq:fidelity}, we have the probability $\frac{1}{8} \cos^2\Phi (1 + \cos^2\Theta_{AB})$ each for $M_{23, 34}$. 
The secret key rate, $R$ in this protocol depends on the error rates $e_X$ (diagonal basis), \; $E_Z$ (rectilinear basis) in a complicated way:
\begin{equation}
R \geq P_{Z}^{1,1} Y_{Z}^{1,1} \left[ 1 - H_2(e_{Z}^{1,1}) \right] - Q_Z f_e(E_Z) H_2(E_Z)
\end{equation}
Here, $i \in \{ Z, X\}$ (basis); $P_i$ is the probability of basis selection, $Y_i$ is the yield (success), $H_2$ is the binary entropy, $Q_i$  is the gain and $f_e$ is the error correction inefﬁciency function. In practice,  protocols implemented using weak coherent sources (WCS), are less prone to frequency mismatch since most CW lasers have narrow bandwidth and easy to match the central frequency, hence overlooked. However, implemented by single photon sources (e.g., heralded spdc sources) has broader bandwidth and may have significant contribution to the error rate. The error parameters $e_X$ and $E_Z$ are calculated directly from the experimental data. The major contribution comes from background noise $(e_b)$, beam splitter asymmetry $(e_a)$, polarization $(e_p)$ and modal/temporal $(e_t)$ mismatch. Since these are derived from the associated fidelity \cite{Xu_2013}, the spectral mismatch error can be quantified as $ e_f  = \dfrac{\sin^2\Theta }{2}$. The total error $e_X$ is then simply the linear combination of all these contributors: $e_X=\sum_i e_i$.

\SetTblrInner{rowsep=2pt}
\begin{table}[H]
\centering
\scalebox{1}{
\begin{tblr}{ |[2pt]Q[c,m ] |[1pt, black] Q[c,m, vio!40] |[1pt, white] Q[c,m, blu!30] |[1pt, white] Q[c,m, olv!30] |[1pt, white] Q[c,m, prp!30] |[2pt]}
\cline[2 pt, black]{-}
 Input & \SetCell[c=4]{c,m,white} {Measurement Probability} & & & \\ \cline[0 pt, white]{2-5}
 State & \textbf{M$_{12}$} & \textbf{M$_{34}$} & \textbf{M$_{23}$} & \textbf{M$_{14}$} \\\cline[2pt,black]{-}
\(\ket{HH}\)         & 0           & 0           & 0           & 0           \\\cline[1pt]{1-5}
\(\ket{HV}\)         & $\frac{1}{4}$ & $\frac{1}{4}$ & $\frac{1}{4}$ & $\frac{1}{4}$ \\\cline[1pt]{1-5}
\(\ket{VH}\)         & $\frac{1}{4}$ & $\frac{1}{4}$ & $\frac{1}{4}$ & $\frac{1}{4}$ \\\cline[1pt]{1-5}
\(\ket{VV}\)         & 0           & 0           & 0           & 0           \\\cline[1pt]{1-5}
\(\ket{DD}\)         & $\frac{1}{4}$ & $\frac{1}{4}$ & 0           & 0           \\\cline[1pt]{1-5}
\(\ket{DA}\)         & 0           & 0           & $\frac{1}{4}$ & $\frac{1}{4}$ \\\cline[1pt]{1-5}
$\ket{AD}$           & 0           & 0           & $\frac{1}{4}$ & $\frac{1}{4}$ \\\cline[1pt]{1-5}
\(\ket{AA}\)         & $\frac{1}{4}$ & $\frac{1}{4}$ & 0           & 0           \\\cline[2pt,black]{-}
\end{tblr}
}
\caption{ Measurement outcome probabilities in setup \ref{fig:MDI-QKD} for various projection operators defined in \ref{eq:M1}, without spectral mismatches.  }
\label{tab:}
\end{table}

Our simulated results illustrate the role that simultaneous misalignment across the modes of photons when interfering at a beam splitter reduces the interference visibility, which in turn affects the security of the protocol by increasing the QBER. Furthermore, our analysis can be extended to identify optimal regimes of operation where visibility is maximized, and consequently, the key rate. Mitigation strategies, such as dynamic polarization control and spectral filtering, can be directly informed by our model, leading to more robust MDI-QKD implementations in real-world environments where source and channel imperfections are unavoidable.

\subsection{Quantum Sensing} 
Quantum sensing applications, particularly those involving interferometric techniques, rely on high-visibility photon interference to achieve enhanced measurement sensitivity. In this context, our multi-photon, multi-mode HOM model provides a useful tool for understanding how mode distinguishability impacts the precision of quantum measurements. For instance, in NOON-state interferometry, which is used to achieve Heisenberg-limited phase estimation, the introduction of polarization or spectral mismatches can degrade the visibility of interference, thereby reducing the sensitivity of the measurement.

As an example, consider the case where Alice and Bob share the multi-photon entangled state \cite{doi:10.1126/science.1104149}
\begin{equation}
    \ket{\Psi}_{AB} \propto \left( \hat{A}^\dagger [\phi_A] \hat{B}^\dagger [\phi_B] \right)^{N-1} \left( \hat{A}^\dagger [\phi_A] + \hat{B}^\dagger [\phi_B] \right) \ket{0}
\end{equation}
where we ignored the polarization of the photons. They send their photons to a 50:50 beam splitter which transforms the input modes according to the relations \eqref{eq:31bs}. One of the output beams acquires an additional phase $\varphi$ which we are interested in measuring. Then the two beams go through a second 50:50 beam splitter which transforms the modes to
\begin{align}\label{eq:31bsa}
    \hat{a} (\omega) &\rightarrow \frac{1}{\sqrt{2}} \left( \hat{a} (\omega) + e^{i\varphi} \hat{b} (\omega) \right), \nonumber\\ 
    \hat{b} (\omega) &\rightarrow \frac{1}{\sqrt{2}} \left( \hat{a} (\omega) - e^{i\varphi} \hat{b} (\omega) \right)
\end{align}
Finally, two detectors measure the intensities of the two outputs, $I_{A} \propto \int d\omega \braket{ \hat{a}^\dagger (\omega) \hat{a} (\omega)}$, $I_{B} \propto \int d\omega \braket{ \hat{b}^\dagger (\omega) \hat{b} (\omega)}$. We are interested in the difference,
\begin{equation}
    \Delta I \propto \int d\omega \braket{ \hat{b}^\dagger (\omega) \hat{b} (\omega) - \hat{a}^\dagger (\omega) \hat{a} (\omega)}
\end{equation}
In terms of the original beams, this is
\begin{align}
    &\int d\omega \; \Big( \braket{\hat{b}^\dagger (\omega) \hat{b} (\omega) - \hat{a}^\dagger (\omega) \hat{a} (\omega)} \cos\varphi \nonumber\\ &+ i \braket{\hat{a}^\dagger (\omega) \hat{b} (\omega) - \hat{b}^\dagger (\omega) \hat{a} (\omega)} \sin\varphi \Big) \nonumber\\ &= - N\cos\Theta_{AB} \sin\varphi
\end{align}
In the ideal case ($\Theta_{AB} =0$), we obtain a sensitivity that scales as $\frac{1}{N}$. With spectral mismatch, the sensitivity scales as $\frac{1}{N\cos\Theta_{AB}}$.

Our model highlights how reducing mode mismatches can lead to improvements in phase sensitivity and noise reduction, which are critical for applications such as gravitational wave detection and atomic clocks. Moreover, the ability to identify specific sources of mode mismatch enables the development of real-time error correction techniques that can dynamically compensate for these imperfections. Such advancements will enhance the robustness of quantum sensors, ensuring that they maintain high sensitivity even in the presence of environmental noise or other operational imperfections. In resource-limited settings, the insights provided by our model will also contribute to more efficient allocation of experimental resources, leading to cost-effective improvements in quantum sensing technologies.

\subsection{Quantum Optical Classifiers}
Quantum optical classifiers, which exploit the quantum interference of photons to perform classification tasks, are highly sensitive to the indistinguishability of the input photons in various degrees of freedom. These classifiers often rely on precise photon interference patterns to make decisions based on input data. Our multi-mode, multi-photon HOM interference model provides a valuable framework for analyzing and improving the performance of quantum optical classifiers in the presence of real-world imperfections, such as spectral and polarization mismatches.

In practical implementations of quantum optical classifiers, photons are sourced from different locations or quantum systems, each with potentially distinct spectral and polarization characteristics. Our model demonstrates how even slight mismatches in these properties can degrade the visibility of quantum interference, leading to reduced classification accuracy. Specifically, our results suggest that controlling for mode distinguishability is crucial for maintaining high classification fidelity in multi-photon interference-based classifiers.

Moreover, the ability to identify and quantify the impact of distinguishability on classifier performance allows for the development of mitigation strategies. Techniques such as adaptive mode-matching, dynamic polarization control, and spectral filtering, informed by our model, can be employed to minimize mode mismatches and optimize the interference visibility. These improvements are critical for scaling quantum optical classifiers to larger, more complex datasets, where maintaining high interference visibility across multiple input modes is essential for reliable classification outcomes.

In addition, quantum optical classifiers may involve interactions between multiple photons in a network of beam splitters and detectors. The insights provided by our model into how mode mismatch affects coincidence probabilities in multi-photon interference experiments can be extended to design more robust classifier architectures. By optimizing the indistinguishability of photons at each interference point, it is possible to enhance the overall accuracy and efficiency of quantum optical classifiers, making them more suitable for practical, large-scale quantum computing tasks.

\begin{table}[ht!]
\centering
\scalebox{0.8}{
\begin{tblr}{ |[2pt]Q[l,m ] |[1pt, white] Q[c,m, red!30] |[1pt, white] Q[c,m, yellow!50] |[1pt, white] Q[c,m, gray9] |[1pt, white] X[c,m, violet7!50] |[2pt]}
\cline[2pt,black]{-}
Object  & $\blacklozenge$ & $\clubsuit$  & $\cdots$ & $\bigstar$ \\ \cline[1pt]{1-5}
Label & $\mathcal{O}_{_0}$ & $\mathcal{O}_1$ & $\cdots$ & $\mathcal{O}_{_{M-1}}$ \\ \cline[1pt]{1}
Binary Target Level  & $y_{_0}$  & $y_{_1}$  & $\cdots$  & $y_{_{M-1}}$ \\ \cline[1pt]{1}
Bias & $b$ & $b$ & $\cdots$ & $b$ \\ \cline[1pt]{1}
\textbf{\textcolor{red}{Training Parameter}} & $\lambda_{1}, \lambda_{2}$ & $\lambda_{1}, \lambda_{2}$ & $\cdots$ & $\lambda_{1}, \lambda_{2}$ \\ \cline[1pt]{1}
{\textbf{\textcolor{red}{Independent  Static Loss}} \\ $C = \alpha_{\lambda_{1,2}}(\mathcal{O}) \;\; \forall \lambda,j$} & C & C & $\cdots$ & C\\ \cline[1pt]{1}
{\textbf{\textcolor{red}{Model prediction}} \\ $f_{\lambda_1, \lambda_2}^{'} = C - 2p(1_a \cap 1_b | \lambda_1, \lambda_2, \mathcal{O})$} & $f_{\lambda_1, \lambda_2}^{'0}$ & $f_{\lambda_1, \lambda_2}^{'1}$ & $\cdots$ & $f_{\lambda_1, \lambda_2}^{^{'M-1}}$ \\ \cline[1pt]{1}
{\textbf{\textcolor{red}{Activation Function}} \\ $F_{b,\lambda_1,\lambda_2}^{'} = \sigma(f^{'} + b)$ } & $F_{b,\lambda_1,\lambda_2}^{'0}$ & $F_{b,\lambda_1 ,\lambda_2}^{'1}$ & $\cdots$ & $F_{b,\lambda_1,\lambda_2}^{^{'M-1}}$ \\\cline[1pt]{1}
{\textbf{\textcolor{red}{Loss Function}} \\  {$H' \!=\! -y\log F' \!-\! (1\!-\!y)\log (1\!-\!F')$} } &  $H^{'}_{_{0}}$ & $H^{'}_{_{1}}$ & $\cdots$  & $H^{'}_{_{M-1}}$ \\
\cline[2pt,black]{-}
\end{tblr}
}
\caption{Modified cross-entropy by including mismatch effect in coincidence probability}
\label{tab:classifier}
\end{table}

Following Ref.\ \cite{roncallo2024quantum}, we consider a beam propagating in the $z$-direction. We will ignore the polarization, but for two-dimensional images we are interested in the transverse components $\bm{k}_\perp = (k_x,k_y)$ of the wave vector. Therefore, we consider creation operators $\hat{a}_{\bm{k}_\perp}^\dagger (\omega)$ obeying commutation relations
\begin{equation}
    [ \hat{a}_{\bm{k}_\perp} (\omega) , \hat{a}_{\bm{k}_\perp'}^\dagger (\omega')] = \delta (\omega - \omega') \delta^2 (\bm{k}_\perp - \bm{k}_\perp')
\end{equation}
and define
\begin{equation}
    \hat{A} [\phi;\phi^\perp] = \int d\omega d^2 k_\perp \phi(\omega) \phi^\perp (\bm{k}_\perp) \hat{a}_{\bm{k}_\perp} (\omega)
\end{equation}
The profiles are normalized as:
\[ \int d\omega |\phi (\omega)|^2 = \int d^2k_\perp |\phi^\perp (\bm{k}_\perp)|^2 = 1 \]
Consider two single-photon states, $\ket{A} = \hat{A}^\dagger [\phi_A;\phi_A^\perp] \ket{0}$ and $\ket{B} = \hat{B}^\dagger [\phi_B;\phi_B^\perp] \ket{0}$. For HOM interference, we send them through a 50:50 beam splitter whose effect is to transform
\begin{equation}
   U_{BS} \ \ : \ \ \begin{array}{lll} \hat{a}_{\bm{k}_\perp} (\omega) &\to &\frac{1}{\sqrt{2}} \Big( \hat{a}_{\bm{k}_\perp} (\omega) + \hat{b}_{\bm{k}_\perp} (\omega) \Big) \ , \\ \hat{b}_{\bm{k}_\perp} (\omega) &\to &\frac{1}{\sqrt{2}} \Big( \hat{b}_{\bm{k}_\perp} (\omega) - \hat{a}_{\bm{k}_\perp} (\omega) \Big) \end{array}
\end{equation}
We place two detectors at the output ports of the beam splitter. If both detectors click, the state is projected onto
\begin{equation}
   \ket{\mathrm{out}} \propto M_0\cdot U_{BS} \ket{A}\otimes \ket{B} \ ,
\end{equation}
where
\begin{equation}
   M_0 = \int d\mu\ \hat{a}_{\bm{k}_\perp}^\dagger (\omega)\hat{b}_{\bm{k}_\perp'}^\dagger (\omega') \ket{0}\bra{0} \hat{a}_{\bm{k}_\perp} (\omega)\hat{b}_{\bm{k}_\perp'} (\omega')
\end{equation}
($d\mu = d\omega d\omega' d^2 k_\perp d^2 k_\perp'$) with probability
\begin{equation}
    p_0 = \bra{B}\otimes\bra{A} U_{BS}\cdot M_0 \cdot U_{BS} \ket{A}\otimes \ket{B}
\end{equation}
It is straightforward to show that
\begin{equation}\label{eq:QOC-copro}
    p_0 = \frac{1}{2} \Big( 1 -  \cos^2 \Theta_{AB} \cos^2 \Theta_{AB}^\perp \Big)
\end{equation}
where the angles measure the overlap between the two single-photon states,
\begin{align}
    \cos\Theta_{AB} &= \left| \int d\omega \phi_A(\omega) \phi_B^* (\omega) \right| \ , \nonumber\\ \cos\Theta_{AB}^\perp  &= \left| \int d^2 k_\perp \phi_A^\perp(\bm{k}_\perp) \phi_B^{\perp *} (\bm{k}_\perp) \right|
\end{align}
With perfect matching, the probability vanishes ($p_0=0$, since $\Theta_{AB} = \Theta_{AB}^\perp=0$). For image processing, we need a finite angle $\Theta_{AB}^\perp$ that maximizes the probability $p_0$. A finite $\Theta_{AB}$ (spectral mismatch) reduces the efficiency of image processing. In the worst case scenario, $\Theta_{AB} = \frac{\pi}{2}$, we obtain maximum 50\% probability without any effort, therefore, no image processing is possible.

In Ref.\ \cite{roncallo2024quantum}, gradient descent was chosen as the iterative optimization algorithm to minimize the loss function for the training procedure. The conditional coincidence probability $p(1_A \cap 1_B | \lambda, \mathcal{O} )$ was used as an argument for both the loss function, $H(y_j, F_{b\lambda})$ (binary cross entropy), and the activation function, $F_{b\lambda}$ (sigmoid) intrinsically and extrinsically, respectively. The descriptions are listed in Table \ref{tab:classifier}.
The sources were assumed to be in single frequency mode. In that case, the mismatch in the momentum yields the coincidence probability $p_0 = \dfrac{1}{2}\sin^2\Theta^\perp_{AB}$, derived from Eq.\ \eqref{eq:QOC-copro} by setting $\Theta_{AB} =0$. 
By tuning the training parameter $\lambda$, one can minimize $\Theta_{AB}^\perp$ for  better visibility. However, if there is an inherent frequency mismatch between the sources, tuning the training parameter $\lambda$ alone may fail (depending on fixed to varying mismatches) to achieve the desired classification result. Spectral mismatches can be removed by optimizing hardware setup (source engineering) or by adding a new trainable parameter, $\lambda \rightarrow (\lambda_1, \lambda_2)$. Table \ref{tab:classifier} shows modified quantities (in bold) if a second training parameter is included.    
If no measure is taken, the lower bound of the effective coincidence probability, $p_0 \in [0, 1/2]$, may elevate from 0 to $\gamma = \dfrac{1}{2}\sin^2\Theta_{AB}$ (a function of the spectral mismatch $\Theta_{AB}$, with $0 \leq \gamma \leq 1/2$) for multi-mode incidences, even though the single training parameter $\lambda$ remains optimized for ideal single-mode condition.

\subsection{Photonic quantum computing}

Linear Optical Quantum Computing (LOQC) leverages single photons, linear optics, and projective measurements for quantum computation. The KLM protocol \cite{Knill2001} enables scalable LOQC by probabilistically inducing photon-photon interactions, circumventing their natural non-interacting behavior. Central to LOQC is the Hong-Ou-Mandel (HOM) effect, which facilitates critical two-photon gates (e.g., CZ and CNOT \cite{Kok_2007}) through constructive/destructive interference, projecting photons into specific computational states.

High-fidelity LOQC gates demand perfect overlap of photonic modes (spectral, temporal, polarization). Imperfect overlap degrades interference visibility, introducing computational errors.

Fusion-Based Quantum Computation (FBQC) \cite{Bartolucci2023} offers an alternative paradigm, constructing large entangled networks via fusion of smaller resource states (e.g., Bell pairs) \cite{bartolucci2021creationentangledphotonicstates}. Unlike traditional methods reliant on direct qubit interactions, FBQC employs projective measurements to fuse cluster states, enabling modular, fault-tolerant gate operations. Fusion success hinges on probabilistic heralding akin to LOQC, with failures mitigated through redundancy.

Both LOQC and FBQC require photon indistinguishability across spectral, spatial, and polarization modes. Imperfect indistinguishability reduces fusion efficiency and gate fidelity, limiting computational scalability \cite{alexander2024manufacturable, rimock2024generalized}.

Consider the two one-dimensional cluster states consisting of $M$ and $N$ qubits, respectively,
\begin{align}
    \ket{C}_A &= \cdots U^{CZ}_{-2,-1} \cdot U^{CZ}_{-1,0} \cdot U^{CZ}_{0,1} \cdot U^{CZ}_{1,2}  \cdots \ket{D}^{\otimes M} \nonumber\\
    \ket{C}_B &= \cdots U^{CZ}_{-2,-1} \cdot U^{CZ}_{-1,0} \cdot U^{CZ}_{0,1} \cdot U^{CZ}_{1,2}  \cdots \ket{D}^{\otimes N}
\end{align}
where $\ket{D} = \frac{1}{\sqrt{2}} (\ket{H} + \ket{V})$ and
\begin{equation}
    U_{i,i+1}^{CZ} = \frac{1}{2} \Big( \mathbb{I} + Z_i + Z_{i+1} - Z_i Z_{i+1} \Big)
\end{equation}
The string of qubits with diagonal polarizations in the first cluster state are created by $A_{D,i}^\dagger \equiv \frac{1}{\sqrt{2}} (A_{H,i}^\dagger + A_{V,i}^\dagger)$, and similarly for the second cluster state, by $B_{D,i}^\dagger \equiv \frac{1}{\sqrt{2}} (B_{H,i}^\dagger + B_{V,i}^\dagger)$, where $i \in \{ \dots, -2,-1,0,1,2, \dots \}$. We can create a two-dimensional cluster state by fusing such one-dimensional cluster states. To this end, we will fuse the qubits at position $0$ in the two cluster states. Before we fuse them, we perform a measurement of the polarization of the qubit in position $1$ in the state $\ket{C}_A$ in the diagonal basis. This projects the state of that qubit to $\ket{D}$ with probability 50\%. We obtain a reduced cluster state with $M-1$ qubits,
\begin{equation}
    \ket{C'}_A = {}_0\braket{D | C}_A = \frac{1}{\sqrt{2}}(\mathbb{I} + Z_{0} Z_2 ) \ket{\chi}_A \ , 
\end{equation}
where
\begin{equation}
    \ket{\chi}_A = \cdots U^{CZ}_{-2,-1} \cdot U^{CZ}_{-1,0} \cdot U^{CZ}_{2,3}  \cdots \ket{D}^{\otimes (M-1)}
\end{equation}
Next, we fuse the two qubits at $0$ positions by taking them through a PBS aligned along the diagonal and detecting the two outputs. We are interested in the case where both detectors click. This will occur with 50\% probability. After going through the PBS, the modes $a_{D,0}^\dagger (\omega)$ and $b_{D,0}^\dagger (\omega)$ exit through the same port, so $a_{D,0}^\dagger (\omega) , b_{D,0}^\dagger (\omega) \to c_{D}^\dagger (\omega)$ and similarly for the other PBS output, $a_{A,0}^\dagger (\omega) , b_{A,0}^\dagger (\omega) \to c_{A}^\dagger (\omega)$. If both detectors click, then the combined state is projected onto the fused state
\begin{equation}    \ket{\mathcal{F}} \propto M_0 \ket{C'}_A \otimes \ket{C}_B \ , 
\end{equation}
where
\begin{equation}
    M_0 = \int d\omega d\omega' \hat{c}_D^\dagger (\omega) \hat{c}_A^\dagger (\omega') \ket{0}\bra{0} \hat{c}_D (\omega) \hat{c}_A (\omega')
\end{equation}
To calculate the fused state $\ket{\mathcal{F}}$, we isolate the mode at position $0$ and write the two cluster states, respectively, as
\begin{align}
    \ket{C'}_A &= \frac{1}{{2}}(\mathbb{I} + Z_{0} Z_2 ) (\ket{H}_0 + Z_{-1} \ket{V}_0) \ket{\chi'}_A \ , \nonumber\\
    \ket{C}_B &= \frac{1}{\sqrt{2}}(\ket{H}_0 + Z_{-1} Z_1 \ket{V}_0 ) \ket{\chi''}_B \ , 
\end{align}
where
\begin{align}
    \ket{\chi'}_A &= \cdots U^{CZ}_{-2,-1} \cdot U^{CZ}_{2,3}  \cdots \ket{D}^{\otimes (M-2)}\nonumber\\
    \ket{\chi''}_B &= \cdots U^{CZ}_{-2,-1} \cdot U^{CZ}_{1,2}  \cdots \ket{D}^{\otimes (N-1)}
\end{align}
In the ideal case, we obtain
\begin{equation}
    \ket{\mathcal{F}} = U^{CZ,A}_{2,-1}\cdot U^{CZ,B}_{2,-1}\cdot U^{CZ,B}_{2,1}  \ket{\chi'}_A \otimes \ket{\chi''}_B \otimes \hat{C}_D^\dagger \hat{C}_A^\dagger \ket{0}
\end{equation}
which is a cluster state in the form of a cross consisting of two one-dimensional cluster states with $M-2$ and $N$ qubits, respectively, sharing the qubit at position $2$ in the cluster state $A$. With spectral mismatch, the cluster state does not decouple from the detectors, and after tracing them out, we are left with a mixed state of fidelity 
\be F = \frac{1}{2} (1 + \cos^2\Theta_{AB}) \ . \ee
It is straightforward to extend the above discussion to include polarization mismatch. It should be noted that such an extension is not needed for 
dual-rail encoding.

\section{Conclusion}\label{sec:6}
In the realm of quantum networking, two-photon interference between independent sources serves as a fundamental building block. However, the practical realization of HOM interference is not without challenges. Realistic imperfections in sources and experimental components, such as beam splitter asymmetries and intensity mismatches, can degrade interference visibility. In this paper, we extend existing works to simultaneously consider the contributions of polarization and spectro-temporal mismatch on HOM visibility, in addition to overall input intensities, beam splitter asymmetries, and detector efficiencies. We aim to provide a comprehensive understanding of multi-photon interference phenomena in HOM setups, particularly in the context of quantum networking. Our investigation opens avenues for linking individual nodes with different quantum resources and capabilities, a crucial step in networking parties via quantum channels with heterogeneous quantum technologies, thereby advancing the development of robust and scalable quantum communication networks.

The implications of our study extend beyond quantum communication and sensing to the broader field of photonic quantum computing. By addressing the challenges of multi-mode distinguishability in multi-photon interference, we further develop the path to large quantum information processing systems. In particular, our findings on mode mismatch and quantum channel effects are crucial for the development of fault-tolerant quantum computers that leverage linear optics. As quantum technologies continue to evolve, the need for precise models and simulations becomes ever more critical. Our work contributes to this effort by providing a detailed analysis of the factors that influence HOM interference and their practical implications for quantum networks, sensing, and computation.

\acknowledgments

This material is based upon work supported by the U.S. Department of Energy, Office of Science, Office of Advanced Scientific Computing Research, through the Quantum Internet to Accelerate Scientific Discovery Program under Field Work Proposal 3ERKJ381. We also acknowledge support by the National Science Foundation under award DGE-2152168.

\bibliography{mp_sources}


\appendix

\section{Superconducting Nanowire Single-Photon Detectors (SNSPDs)} \label{app:4}

Superconducting nanowire single-photon detectors (SNSPDs) are state-of-the-art devices renowned for their high detection efficiency, low timing jitter, and low dark count rates. However, the performance of SNSPDs can be significantly influenced by the polarization of the incoming photons. Understanding the polarization sensitivity and efficiency of these detectors is crucial for their application in quantum optics and quantum information science, where precise control and measurement of single photons are essential.

The polarization sensitivity of SNSPDs arises from the anisotropic nature of the nanowire's material and geometry. Typically, these detectors are made from superconducting materials like niobium nitride (NbN) or tungsten silicide (WSi), patterned into thin, meandering nanowires. The absorption of photons in the nanowire depends on the orientation of the photon's electric field relative to the nanowire's axis. Photons polarized parallel to the nanowire are more efficiently absorbed than those polarized perpendicularly. This anisotropy stems from the fact that the superconducting current density and the electric field of the photon interact differently depending on their relative orientations \cite{Natarajan_2012}.

Empirical studies have shown that polarization sensitivity can lead to variations in detection efficiency by as much as 20–30\% between different polarization states \cite{Marsili2013}. To mitigate this effect, SNSPDs can be engineered with polarization-independent designs. One approach involves using nanowire arrays oriented in different directions or implementing a circularly polarized light configuration to average out the polarization dependence \cite{Verma_2014}. Additionally, integrating optical components like quarter-wave plates can convert linearly polarized light into circularly polarized light before it reaches the detector, thus reducing polarization sensitivity \cite{Sahin2013}.

The detection efficiency of SNSPDs, often referred to as the system detection efficiency (SDE), is defined as the probability that an incident photon is successfully detected and registered as an electrical signal. It encompasses several factors: the intrinsic efficiency of the nanowire material, the optical coupling efficiency, and the electronic readout efficiency. Intrinsic efficiency is primarily determined by the probability that an absorbed photon will generate a detectable electrical pulse. This probability is high due to the rapid transition of the superconducting nanowire to a resistive state upon photon absorption \cite{kerman_2007}. The optical coupling efficiency is influenced by how well the incoming photons are directed onto the active area of the nanowire. Techniques such as integrating optical cavities or using anti-reflection coatings can enhance this efficiency \cite{Sprengers_2011}. In practical applications, the SDE of SNSPDs can exceed 90\% at specific wavelengths, notably in the telecom range around 1550 nm, making them highly suitable for fiber-optic communication and quantum key distribution. However, the efficiency can drop at other wavelengths or under less-than-ideal coupling conditions. Therefore, optimizing the optical setup, including fiber alignment and focusing optics, is critical to maintaining high detection efficiency.

An arbitrary polarization state can be decomposed into its horizontal and vertical components. The polarization state of a photon with polarization vector $\hat{\epsilon}$ can be expressed as
\begin{equation}
    |\hat{\epsilon}\rangle = \alpha |H\rangle + \beta |V\rangle
\end{equation}
where $|H\rangle$ and $|V\rangle$ represent the horizontal and vertical polarization states, respectively, and $\alpha$ and $\beta$ are the complex probability amplitudes given by $\alpha = \hat{\epsilon} \cdot \hat{\epsilon}_H$ and $\beta = \hat{\epsilon} \cdot \hat{\epsilon}_V$. The condition $|\alpha|^2 + |\beta|^2 = 1$ ensures normalization of the state. The detection efficiency $\eta$ for a photon in state $|\hat{\epsilon}\rangle$ can be calculated as a weighted average of the efficiencies for $|H\rangle$ and $|V\rangle$:
\begin{equation}
    \eta = |\alpha|^2 \eta_H + |\beta|^2 \eta_V
\end{equation}
where $\eta_H$ and $\eta_V$ are the detection efficiencies for horizontally and vertically polarized photons, respectively. Here, $\eta$ represents the probability that a photon in state $|\hat{\epsilon}\rangle$ is detected, with $\eta = 1$ indicating perfect detection.

Given $m$ photons with polarization $\hat{\epsilon}_A$ and $n$ photons with polarization $\hat{\epsilon}_B$, the probability of not detecting any photons is: 
\begin{equation}
    P(\text{no clicks}) = (1 - \eta_A)^m (1 - \eta_B)^n 
\end{equation}
Detecting at least one photon is the complement of this event, with probability of a detector click given by: 
\begin{align}\label{eq:def}
    \Delta(\hat{\epsilon}_A, \hat{\epsilon}_B, m, n) &\equiv P(\text{at least one click}) \nonumber\\ &= 1 - (1 - \eta_A)^m (1 - \eta_B)^n
\end{align}

\section{Polarization and Spectro-temporal Distinguishability}
\label{app:1}

Here, we provide a detailed mathematical derivation of the detection probabilities for multi-photon and coherent states after passing through a beam splitter, which are needed for the coincidence probability given in the main text.

Consider an input state to the beam splitter consisting of $m$ photons from source $A$ and $n$ photons from source $B$.

After passing through a beam splitter of transmittance $t$ and reflectivity $r$, it undergoes spatial mixing of creation operators resulting in the state given by Eq.\ \eqref{eq:6} in the main text.

By applying the binomial theorem, the output state can be written as
\begin{widetext}
\begin{equation}
    \begin{split}
        \ket{\text{out}} = \frac{1}{\sqrt{m!n!}} &\sum_{k=0}^m \sum_{l=0}^n (-1)^l\binom{m}{k} \binom{n}{l} t^{m+l-k} r^{n+k-l}  \\
        &\times \bigg{(} \hat{A}^\dagger [\hat{\epsilon}_A, \phi_A] \bigg{)}^{m-k} \bigg{(} \hat{A}^\dagger [\hat{\epsilon}_B, \phi_B] \bigg{)}^{n-l} \bigg{(} \hat{B}^\dagger [\hat{\epsilon}_A, \phi_A] \bigg{)}^k \bigg{(} \hat{B}^\dagger [\hat{\epsilon}_B, \phi_B] \bigg{)}^l \ket{0}_A\otimes \ket{0}_B
    \end{split}
\end{equation}
    
\end{widetext}
where $\hat{A}^\dagger [\hat{\epsilon}, \phi]$ is defined in Eq.\ \eqref{eq:2a} and similarly for $\hat{B}^\dagger$.

We are interested in determining the probability of coincidence detection for this state given by Eq.\ \eqref{eq:7} in the main text. To calculate $P_{m,n}^\text{det}(m+n,0)$, we introduce the projector for detecting all photons at output port $A$,
\begin{equation}
\begin{split}
    \hat{P}^A_{m+n} &= \frac{1}{(m+n)!}\sum_{i_1, \cdots, i_{m+n}} \int d\omega_1 \cdots \int d\omega_{m+n} \\
    &\hat{a}_{i_1}^\dagger({\omega}_1)  \cdots \hat{a}_{i_{m+n}}^\dagger({\omega}_{m+n}) \ket{0} \bra{0} \hat{a}_{i_1}({\omega}_1)  \cdots \hat{a}_{i_{m+n}}({\omega}_{m+n})
\end{split}
\end{equation}

where the detector is assumed ambivalent to polarization direction and spectral properties. In general, detectors are only sensitive to a range of frequencies, but under the assumption that the bandwidth of the sourced photons is much narrower than that of the detector being used, we can ignore this constraint. Polarization sensitivity can be included as outlined in Appendix \ref{app:4}. 
Evaluating the projector at the output state, after some algebra we obtain the detection probability
\begin{widetext}
\begin{equation}
    \begin{split}
       P_{m,n}^{\text{det}}(m+n, 0) &= \bra{\text{out}}\hat{P}^A_{m+n} \ket{\text{out}} \\ &= \frac{t^{2m} r^{2n}}{m!n!(m+n)!}  \sum_{i_1, \cdots, i_{m+n}} \int d\omega_1 \cdots \int d\omega_{m+n} \left| \bra{0} \bigg{(} \hat{A}[\hat{\epsilon}_B, \phi_B] \bigg{)}^n \bigg{(} \hat{A}^\dagger [\hat{\epsilon}_A, \phi_A] \bigg{)}^m \hat{a}_{i_1}^\dagger({\omega}_1)  \cdots \hat{a}_{i_{m+n}}^\dagger({\omega}_{m+n}) \ket{0} \right|^2 
    \end{split}
\end{equation}
    
\end{widetext}
After a straightforward, albeit tedious, calculation of inner products, we arrive at the expression \eqref{eq:7a} for the detection probability in the main text. The expression for the probability of detecting all photons at the other detector, $P_{m,n}^{\text{det}}(0,m+n)$ is readily deduced from $P_{m,n}^{\text{det}}(m+n, 0)$ by the symmetry of the problem.

Having obtained an expression for detection, we can add considerations of polarization-dependent detector efficiency by inserting the respective detector efficiency functions, $\Delta_A(p_a, p_b, m, n)$ and $\Delta_B(p_a, p_b, m, n)$ (see Appendix \ref{app:4} for details). Then noting $T=t^2$ and $R=r^2$, we deduce the coincidence probability given by Eq.\ \eqref{eq:9} of the main text. 


Turning to coherent states, the coincidence probability for two coherent sources given by Eq.\ \eqref{eq:pmu} reads 
\begin{equation}
        P^{Co}_{\text{total}} = e^{-\mu_A-\mu_B} \sum_{m,n}  \frac{\mu_A^m \mu_B^n}{m!n!} P^{Co}_{m,n}
\end{equation}
Using the result \eqref{eq:9} for a beam splitter of transmittance $T$ and reflectivity $R$, and arbitrary separable spectral photons, we obtain
\begin{widetext}
\begin{equation}
    \begin{split}
         P^{\text{Co}}_{\text{total}} &= e^{-\mu_A-\mu_B}
         \sum_{m+n\geq 0} \frac{\mu_A^m \mu_B^n}{m!n!}\Delta_A\Delta_B \\ 
         &- e^{-\mu_A-\mu_B} \sum_{m+n\geq 0} \frac{\mu_A^m \mu_B^n}{m!n!}\bigg{[}\big{(}R^mT^n\Delta_B+T^mR^n\Delta_A\big{)} \sum_{j=0}^{\text{min}(m,n)}\binom{m}{j}\binom{n}{j}\cos^{2j}\Phi  \cos^{2j}\Theta\bigg{]} 
\end{split}
\end{equation}
    
\end{widetext}
where (see Appendix \ref{app:4} for a derivation)

\begin{equation}
    \begin{split}
        &\Delta_A(\hat{\epsilon}_A, \hat{\epsilon}_B, m, n) = 1-(1-\eta_A)^m(1-\eta_B)^n \\
        &\Delta_B(\hat{\epsilon}_A, \hat{\epsilon}_B, m, n) = 1-(1-\eta_A')^m(1-\eta_B')^n
    \end{split}
\end{equation}

We define four coefficients to simplify the expression,
\begin{equation}
    \begin{split}
        &\mathcal{A} = \mu_A R (1-\eta_A') \\
        &\mathcal{B} = \mu_B T (1-\eta_B') \\
        &\mathcal{C} = \mu_A T (1-\eta_A) \\
        &\mathcal{D} = \mu_B R (1-\eta_B)
    \end{split}
\end{equation}
After some tedious algebra, we arrive at the expression for the total coincidence probability

\begin{widetext}
\begin{equation}\label{eq:B5}
\begin{split}
P_{\text{total}}^{\text{Co}}     &= 1 - e^{-\mu_A \eta_A' - \mu_B \eta_B'} - e^{-\mu_A \eta_A - \mu_B \eta_B} + e^{\mu_A (\eta_A \eta_A' - \eta_A - \eta_A') + \mu_B(\eta_B \eta_B' - \eta_B - \eta_B')} \\
         &- e^{-\mu_A-\mu_B} (e^{\mu_A R + \mu_B T}+e^{\mu_A T + \mu_B R})\text{I}_0(2\sqrt{\mu_A \mu_B R T} \cos\Phi \cos\Theta) \\
         &+ e^{-\mu_A-\mu_B + \mathcal{A} + \mathcal{B}} \text{I}_0(2\sqrt{\mathcal{A}\mathcal{B}} \cos\Phi \cos\Theta) + e^{-\mu_A-\mu_B + \mathcal{C} + \mathcal{D}} \text{I}_0(2\sqrt{\mathcal{C}\mathcal{D}} \cos\Phi \cos\Theta)
    \end{split}
\end{equation}
    
\end{widetext}
where $I_0$ is the modified Bessel function of the first kind with order $0$. Then, assuming ideal detectors, letting $R=T=50\%$, and $\mu_A=\mu_B=\mu$, we recover Eq.\ \eqref{eq:coh_coin} of the main text.


\section{Common Spectral Envelopes for Quantum-Sourced Photons} \label{app:3}
In quantum optics, the characterization of the spectral properties of photons is crucial, particularly in multi-mode, multi-photon interference phenomena such as Hong-Ou-Mandel (HOM) interference. Here we provide more detailed descriptions of several common spectral envelopes for quantum-sourced photons, normalized such that the integral of the spectral envelope over all frequencies equals one. Understanding these spectral profiles is essential for ensuring robust multi-photon interference and networking different quantum sources.

Each of the following spectral envelopes is normalized to satisfy the condition:
\begin{equation}\label{eq:phi}
    \int |\phi(\omega)|^2 d\omega = 1. 
\end{equation} 

\subsection{Gaussian Spectral Envelope}

The Gaussian spectral envelope is widely used due to its mathematical simplicity and its natural occurrence in many physical processes, such as parametric down-conversion and certain types of laser emission. It is given by:
\begin{equation}
    \phi(w) = \left( \frac{1}{\sigma \sqrt{2\pi}} \right)^{1/2} \exp\left( -\frac{(\omega -\omega_0)^2}{4\sigma^2} \right),
\end{equation}  
where $\omega_0$ is the central frequency and $\sigma$ is the spectral width (standard deviation). This profile is symmetric around $\omega_0$. Gaussian spectral envelopes are often utilized in quantum key distribution (QKD) systems and HOM interference experiments to maximize photon indistinguishability.

\subsection{Sinc Spectral Envelope}

The sinc function arises naturally in scenarios involving rectangular time apertures and is common in time-bin encoding and pulsed quantum systems. It is given by:
\begin{equation}
\phi(\omega) = \left( \frac{T}{2\pi} \right)^{1/2} \frac{\sin\left( T(\omega- \omega_0)/2 \right)}{(\omega -\omega_0)/2}, 
\end{equation}
where  $T$ is the time duration of the photon and $\omega_0$ is the central frequency. This profile features a central peak at $\omega_0$ and decaying side lobes. The sinc spectral profile is particularly relevant in photon sources based on time-bin encoding, where precise timing information is crucial.

\subsection{Lorentzian Spectral Envelope}

Common in resonant systems and spontaneous emission processes, the Lorentzian spectral profile arises from sources such as single-photon emitters based on quantum dots or certain atomic transitions. It is given by:
\begin{equation}
\phi(\omega) = \left( \frac{\gamma}{2\pi} \right)^{1/2} \frac{1}{(\omega -\omega_0)^2 + (\gamma/2)^2}, 
\end{equation}
where $\gamma$ is the full width at half maximum (FWHM) and $\omega_0$ is the central frequency. This profile decays more slowly in the wings compared to a Gaussian profile. Lorentzian envelopes are used to describe photons emitted from systems where the spectral linewidth is dominated by natural or radiative broadening, such as in quantum dot and certain atomic systems.

\subsection{Sech Hyperbolic Spectral Envelope}

The hyperbolic secant spectral profile arises in specific coherent light-matter interactions, such as those involving solitons in nonlinear optical fibers. It is given by:
\begin{equation}
\phi(\omega ) = \left( \frac{1}{2\pi} \right)^{1/2} \frac{1}{\cosh((\omega - \omega_0)/\sigma)}, 
\end{equation}
where $\sigma$ determines the width of the spectral profile and $\omega_0$ is the central frequency. This profile features broader peaks and tails compared to a Gaussian profile. The sech spectral profile is utilized in experiments involving solitons and other coherent structures in nonlinear media.

Altogether, these spectral envelopes represent a range of photon spectral properties encountered in quantum optics. For multi-mode, multi-photon HOM interference experiments, understanding and controlling these spectral profiles is crucial. Variations in these profiles can significantly impact the indistinguishability and interference visibility of photons from different sources. The ability to match or tailor these spectral properties is essential for networking users with different quantum sources, enabling robust and high-fidelity quantum communication and computation protocols.


\section{Quantum Channels}
\label{app:Quant_channels}

In the context of quantum information, a quantum channel represents the evolution of a quantum state due to noise or other external factors. Here, we describe three types of quantum channels: amplitude damping, depolarizing, and spectral broadening.

\subsection{Amplitude Damping Channel}

The amplitude damping channel models the loss of energy from a quantum system. For a photon number state, this process can be parameterized by $\gamma$, representing the probability of a photon being lost. The channel acts on the photon number state as follows:
\begin{equation}
    \mathcal{E}_m(\rho) = \sum_k E_k \rho E_k^{\dagger},
\end{equation}
where the Kraus operators \(E_k\) are given by
\begin{equation}
    E_k = \sum_{n} \sqrt{\binom{n}{k}} \sqrt{(1-\gamma)^{n-k} \gamma^k} \ket{n-k}\bra{n}.
\end{equation}

\subsection{Depolarizing Channel}

 The depolarizing channel introduces random noise that depolarizes the quantum state. The Kraus operators for the depolarizing channel are 
 \begin{equation}
     K_0 = \sqrt{1 - p} I, \ K_1 = \sqrt{\frac{p}{3}} X, \ K_2 = \sqrt{\frac{p}{3}} Y, \ K_3 = \sqrt{\frac{p}{3}} Z,
 \end{equation} 
 where $p$ is the depolarizing probability and $X, Y,Z$ are the Pauli matrices acting on the polarization qubit. For a polarization state $\alpha |H\rangle + \beta |V\rangle$, the depolarizing channel acts as
 \begin{equation}
     \mathcal{E}_{\text{DP}}(\rho) = (1 - p) \rho + \frac{p}{3} (X \rho X + Y \rho Y + Z \rho Z),
 \end{equation}
 where $\rho$ is the density matrix of the polarization state.

\subsection{Gaussian Spectral Broadening Channel}

Spectral broadening accounts for changes in the spectral profile of the photon. For Gaussian spectral broadening, this process can be parameterized by a broadening factor \(\xi\). The channel's effect on the spectral component of the state can be modeled as:

\[
\mathcal{E}_{\phi}(\rho) = \int d\omega \, g(\omega) \rho g(\omega)^{\dagger},
\]
where \(g(\omega)\) is a Gaussian function defined as:

\[
g(\omega) = \left( \frac{1}{\sigma \sqrt{2\pi}} \right)^{1/2} \exp\left( -\frac{(w-w_0)^2}{4\sigma^2} \right),
\]
with \(\sigma = \xi \sigma_0\) representing the broadened spectral width, \(\omega_0\) the central frequency, and \(\sigma_0\) the initial spectral width.

\subsection{Combined Action of Quantum Channels}
When considering the combined effect of amplitude damping, depolarizing, and spectral broadening channels on an input pure \(m\)-photon state with polarization and a continuum of spectral modes, the overall channel can be expressed as:

\[
\mathcal{E}(\rho) = \mathcal{E}_m(\rho_m) \otimes \mathcal{E}_{\hat{\epsilon}}(\rho_{\hat{\epsilon}}) \otimes \mathcal{E}_{\phi}(\rho_{\phi}).
\]

Suppose Alice prepares an initial pure state \(\rho_A\) described by:

\[
\rho_A = \rho_m \otimes \rho_{\hat{\epsilon}} \otimes \rho_{\phi},
\]

where \(\rho_m\) represents the photon number state, \(\rho_{\hat{\epsilon}}\) the polarization state, and \(\rho_{\phi}\) the spectral state. After passing through the quantum channels, the state transforms as:

\[
\mathcal{E}(\rho_A) = \mathcal{E}_m(\rho_m) \otimes \mathcal{E}_{\hat{\epsilon}}(\rho_{\hat{\epsilon}}) \otimes \mathcal{E}_{\phi}(\rho_{\phi}).
\]

This indicates that the overall transformation can be seen as a tensor product of the individual transformations applied to each degree of freedom.

For instance, an \(m\)-photon state \(\ket{m}\) undergoing amplitude damping becomes a mixed state with probabilities redistributed among states with fewer photons. Simultaneously, the polarization state \(\rho_{\hat{\epsilon}}\) may experience bit-phase flips due to depolarizing noise. Finally, the spectral state \(\rho_{\phi}\) is broadened according to the broadening function \(g(\omega)\). This demonstrates the interplay between different types of noise and how they collectively affect the state before it reaches the beam splitter for HOM interference. This framework allows for the analysis of the impact of different quantum channels on photon number states with polarization and spectral modes. By modeling amplitude damping, depolarizing, and spectral broadening channels using Kraus operators, we can study how these processes affect the coherence and interference properties of the photons in various experimental scenarios.

\end{document}